\documentclass[twocolumn]{aastex62}
\usepackage{CJK}
\usepackage{amsmath,scalerel}
\usepackage{natbib}
\bibpunct{(}{)}{;}{a}{}{,}
\usepackage{multirow}
\usepackage{graphics}
\usepackage{threeparttable}
\usepackage[varg]{txfonts}
\usepackage{xcolor}
\usepackage{bm}
\hypersetup{colorlinks,linkcolor={blue},citecolor={blue},urlcolor={blue}}
\newcommand{\Msunh}{\>h^{-1}\rm M_\odot}

\newcommand{\Mpch}{\>h^{-1}{\rm {Mpc}}}
\newcommand{\kpch}{\>h^{-1}{\rm {kpc}}}
\newcommand{\kpc}{{\rm \ {kpc}}}
\newcommand{\kms}{\>{\rm km}\,{\rm s}^{-1}}

\newcommand{\magarcsec}{ \ {\rm mag} \ {\rm arcsec^{-2}}}
\def\gtsima{$\; \buildrel > \over \sim \;$}
\def\ltsima{$\; \buildrel < \over \sim \;$}
\def\gta{\lower.7ex\hbox{\gtsima}}
\def\lta{\lower.7ex\hbox{\ltsima}}

\reportnum{}
\received{}
\revised{}
\accepted{}
\published{}
\submitjournal{ApJ}

\shorttitle{satellites distribution and redshift evolution}
\shortauthors{Tang et al.}

\begin{document}
\begin{CJK*}{UTF8}{gkai}
%-----------%
\title{Satellite Alignment: III. Satellite Galaxies Spatial Distribution and their Dependence on Redshift with A Novel Galaxy Finder }
\author{Lin Tang (唐林)}
\affiliation{School of Physics and Astronomy, Sun Yat-sen University,
DaXue Road 2, 519082, Zhuhai, China}
\author{Weipeng Lin (林伟鹏)}
\affiliation{School of Physics and Astronomy, Sun Yat-sen University,
DaXue Road 2, 519082, Zhuhai, China}
\author{Yang Wang  (汪洋)}
\affiliation{School of Physics and Astronomy, Sun Yat-sen University,
DaXue Road 2, 519082, Zhuhai, China}

\correspondingauthor{Lin Tang (唐林), Weipeng Lin (林伟鹏)}
\email{E-mail: tanglin23@mail.sysu.edu.cn, linweip5@mail.sysu.edu.cn}
%-----------%
%--Abstract--%
\begin{abstract}
After extensively explored, broad agreement between observations and theories has been reached that satellites are preferentially aligned with major axes of their host centrals.
There are still some issues unsolved on this topic.
In this paper, we present studies on satellite spatial distribution. 
To fairly compare with observations, we develop a novel galaxy finder and reconstruction algorithm in hydrodynamical simulation, which is based on the projected mock image, taking into account the full consideration of the point spread function, pixel size, surface brightness limit, resolution and redshift dimming effects. 
With galaxy samples constructed using such an algorithm, the satellite alignment is examined by comparing with observational results.
It is found that the observational alignment can be reproduced for red galaxies, which dominate the sample in this study, but not for blue galaxies.
Satellites' radial distribution is also investigated.
It exhibits that outer satellites within host halos show stronger alignment signal than satellites in the inner regions, especially for red satellites, which is in contrast with previous studies.
The disagreement is mainly due to extra galaxies identified by our new galaxy finder, which are mainly located in the inner region of host halos.
Our study illustrates that at lower redshift, the alignment strength becomes stronger, while radial distribution curve becomes flatter.
This suggests differences in the evolution of the angular distribution between satellites residing in the inner and outer halos, and implies that the post-infall evolution reduces the original alignment signal, that the impact decreases for satellites with later infall times.

\end{abstract}
%-----------%
%--keyword--%
%-------------%
%--Introduction--%
\section{Introduction}\label{sec:introduction}
Modern observational results suggest the formation of cosmic structure as hierarchical clustering, where small halos form first and subsequently merge to form bigger ones. 
Although current cosmology based on $\Lambda$CDM model is successful in explaining large cosmic structure \cite[e.g.,][]{Bahcall1999},
A lot of research works have found that there are some serious discrepancies between observations and simulations, such as the core-cusp problem, missing baryons, too-big-too-fail problems, and etc \cite[e.g.,][]{Klypin1999, Maller&Bullock2004, Walker&Penarrubia2011, Boylan-Kolchin2011}.
Those tensions mainly occur in small scales, where the structure formation is dominated by nonlinear astrophysical processes \cite[e.g.,][]{Bertschinger1998, Dolag2008, Kuhlen2012}.
Those aforementioned problem can be partly solved by the modified baryonic model in galaxy formation\cite[e.g.,][]{Bullock&Boylan-Kolchin2017}.
The spatial distribution of satellite galaxies is one of the most important characteristics of small-scale structure, which is associated to the galaxy dynamics and mass distribution \cite[e.g.,][]{Knebe2004,Sales2007}.
Accurate prediction on satellite spatial distribution can provide important clues for structure formation in small scales and may help to solve the tension in some degree.
\par
The study on satellite spatial distribution has a long history.
There was no converged conclusions in early observational studies.
For example, \cite{Holmberg1969} studied the satellites of local galaxies, and found that the satellite galaxies distribute peculiarly to the disk of central galaxies. 
In another word, most of them align along the minor axis of central galaxy, named as Holmberg effect.
However, \cite{Sastry1968} found that there is a strong tendency for the distribution of galaxies to be oriented along the major axes of the cD galaxies (centrals of clusters).
Benefited by the development of large galaxy surveys, such as 2degree Field Galaxy Redshift Survey (2dFGRS), observations has manifested statistical evidence of the Holmberg effect \cite[]{Sales&Lambas2004}, but only for very specific subsample with radial velocity relative to the primary $|\Delta v|<160\kms$.
\par
With a more complete sample and no specific selection criteria, Sloan Digital Sky Survey (SDSS) has shown that the satellites are preferentially distributed along the major axes of centrals \cite[][hereafter Y06]{Yang2006}.
Y06 used galaxy group catalogue of  \cite{Weinmann2006}, which is constructed based on New York University Value-Added Galaxy Catalogue \cite[]{Blanton2005} with galaxy group finder developed by \cite{Yang2005}. This group finder links galaxies into groups by friends-of-friends (FoF) algorithm \cite[]{Davis1985}. 
Then it estimates the host halo mass and viral radius according to the group member galaxies and kick off galaxies out of viral radius. 
This process is repeat iteratively until the galaxy group catalogue converges. 
In each galaxy group, the brightest member is treated as the central galaxy, while other members are treated as satellite galaxies.
In addition, Y06 compared with previous studies and found that the inconsistency is mainly caused by the small sample size and misinterpretation of the position angle.
The phenomenon in Y06 is widely confirmed by following investigations \cite[e.g.,][]{Faltenbacher2007, Bailin2008, Agustsson&Brainerd2010}.
\par
The previous observational studies have investigated that the strength of satellite alignment depends on the galaxy properties, such as color of centrals and satellites\cite[e.g.,][]{Azzaro2007}, radius between with centrals \cite[e.g.,][]{Brainerd2005,Brainerd&Yamamoto2019} and surrounding environment \cite[e.g.,][]{Zhang2015,Wang2018}.
The relationship between satellites spatial distribution and galaxy properties indicates that satellite alignment is connected with galaxy formation and evolution, and can be a tracer of the small scale cosmic structure.
Many theoretical works claimed that satellite alignment can be reproduced in CDM model and interpreted by the non-spherical nature of dark matter halos, in which satellites are aligned with the major axes of host centrals \cite[e.g.,][]{Kang2005a, Agustsson&Brainerd2010, Wang2014a}.
\par
However, the main discrepancy between observational and  theoretical studies is that the alignment strength of observations is commonly weaker than the prediction of simulations \cite[e.g.,][]{Kang2007, Faltenbacher2008, Bett2010,Bahl&Baumgardt2014} and the dependence on galaxy properties is poorly reproduced \cite[e.g.,][]{Agustsson&Brainerd2010}. 
\par
Many works attempted to solve those problems. 
\cite{Kang2007} studied the galaxy alignment using N-body simulation which includes a semi-analytical model for galaxy formation, and compared the results with that in Y06.
They stated that the galaxy catalogue in Y06 is impacted by interloped (i.e., nearby galaxies identified as satellites by the galaxy finder).
Furthermore, they argued that the galaxy catalogue is significantly incomplete, caused by two main selection effects: the observational apparent magnitude limit of $m_r<22.2 \ \rm magnitude$ for galaxies, and the group finder criterion of $M>10^{12}h^{-1}M_{\odot}$ for their host halos.
Those effects reduce the galaxy alignment signal and produce an artificial dependence of alignment strength on the color of the central galaxy.
\cite{Wang2014a} (paper I) used a N-body simulation to explore over-prediction of satellite alignment and its dependence on the mass, formation time of host halos, and accretion time of subhalos. 
The central galaxy shares the same shape as the inner region of host halos, and subhalos are used to trace the positions of satellite galaxies. 
This work can reproduce the satellites spatial distribution (represented by probability function of distribution angles) using inner halo shape, but shows no dependence of alignment on the color of satellites caused by the limitation of pure N-body simulations. 
\cite{Dong2014} (Paper II, hereafter D14) reproduced the observational alignment signal and the color dependence well utilizing a hydrodynamical cosmological simulation. 
They also found that satellite alignment depends strongly on satellite metallicity. 
But the color dependence has low confidence level caused by a lack of AGN feedback in simulation.
\par
\cite{Brainerd&Yamamoto2019} investigated the locations of luminous satellite galaxies using the hydrodynamical simulation Illustris which includes complete galaxy formation model and found that the misalignment between mass and luminosity can affect the anisotropy of satellites distribution.
Moreover, they found that the anisotropy of the satellite distribution decreases with three dimensional distance between satellites and hosts, which is completely in contrast with previous studies \cite[e.g.,][]{Yang2006,Dong2014,Wang2018}.
\par
Moreover, the redshift evolution of galaxy alignment has been poorly studied.
\cite{Donoso2006} used the SDSS DR4 to study the alignment of luminous red galaxies at $z\sim0.5$.
The result is similar as in the local universe.
\cite{Wang2010} created a high-redshift ($0.4<z<1.0$) group catalog out of a spectroscopic sample of galaxies in the GOODS fields and studied the distribution of satellite galaxies. 
They found that there is no significant difference between the alignment strength in high-z and local groups for the total samples, but the generality of this conclusion could have been limited by the small sample size. 
In their discussion, they argue that a weaker alignment signal is expected at higher redshifts.
\cite{Samuroff2019} tested the satellite anisotropy using MASSIVEBLACK-II simulation,
and it was found that there is no evidence of coherent evolution for galaxy intrinsic alignment with redshift, but with large scatter in small scales.
\par
The galaxy is usually defined as a group of bounding particles in simulations. 
Apparently,  it is very different to define bounding groups in observations where surface brightness is in consideration. 
Therefore, when making comparisons, inconsistency in methods will influence the analysis of difference between simulation results and observations, and eventually reduce the reliability of  conclusions.
In our previous paper \cite[]{Tang2018}, we argued that the general trend of intra-cluster light (ICL) redshift evolution in observational results agrees with our theoretical predictions, using the mocking projected image and similar observational parameters in simulation.
we emphasize the importance of using the same definition when observational results are compared with theoretical predictions.
\par
In this paper, we will re-examine the satellite spatial distribution and its dependence on redshifts, using a novel method which mimics observation. 
We develop a novel galaxy finder and reconstruction algorithm in hydrodynamical simulation, which is based on the projected mock image modified with point spread function (PSF), pixel size, surface brightness limit, resolution and redshift dimming effects. 
In such a way, the comparison between observations and theories will be much fairer than previous studies.
\par
The paper is organized as follows. 
In Section~\ref{sec_method}, we briefly describe the simulation we used, how we determine the mock galaxies and galaxy sample with a set of selection criteria. 
In Section~\ref{sec:results}, we show the satellites spatial distribution, including the results of alignment signal, dependence on radii of dark matter halos and redshifts. 
We summarize and briefly discuss our results in Section~\ref{sec:conclusions}.
%-----------%
%--Method---%
\section{Methodology}\label{sec_method}
%--simulation--%
\subsection{simulation}\label{sec_simulation}
The cosmological simulation was run with GADGET-2 code \cite[]{Springel2005a}, which is the same simulation used in previous works \cite[]{Dong2014, Tang2018}.
It run in a $\Lambda$CDM universe, with $\Omega_m=0.268$, $\Omega_{\Lambda}=0.732$,  $\sigma_8=0.85$, $h=0.71$ and $512^3$ DM and $512^3$ gas particles in a cubic box with side length of $100 \Mpch$.
The Plummer softening length is $4.5\kpc$, and each dark matter and gas particle has a mass of  $4.62\times10^8 \Msunh$ and $9.20\times10^7 \Msunh$.
Gas particle can be turned into two star particles later on.
The simulation includes gas cooling, star formation, SN feedback, but no AGN feedback. Dark matter halos are defined by the FoF algorithm with a linking length of 0.2 times the mean particle separation. 
Each star particle of the FoF group is treated as a simple stellar population (SSP) with age, metallicity, and mass given by the corresponding particle’s properties in the simulation.
The FoF algorithm agrees remarkably well on fundamental properties of dark matter halos, comparing with other halo finders using high resolution cosmological simulations \cite[e.g.,][]{Knebe2011}.
Only halos with a minimum particle number of 60 will be selected for later analysis.
Readers interested in detailed description of the simulation are referred to \cite{Springel&Hernquist2003} and \cite{Lin2006}.
%----------------%
%--mock galaxy--%%
\subsection{mock observation of galaxy}\label{sec_mack_glalaxy}
%------------------------%
%--Figure_Luminosity_profile--%
\begin{figure}
  \centering  
    \includegraphics[width=0.45\textwidth]{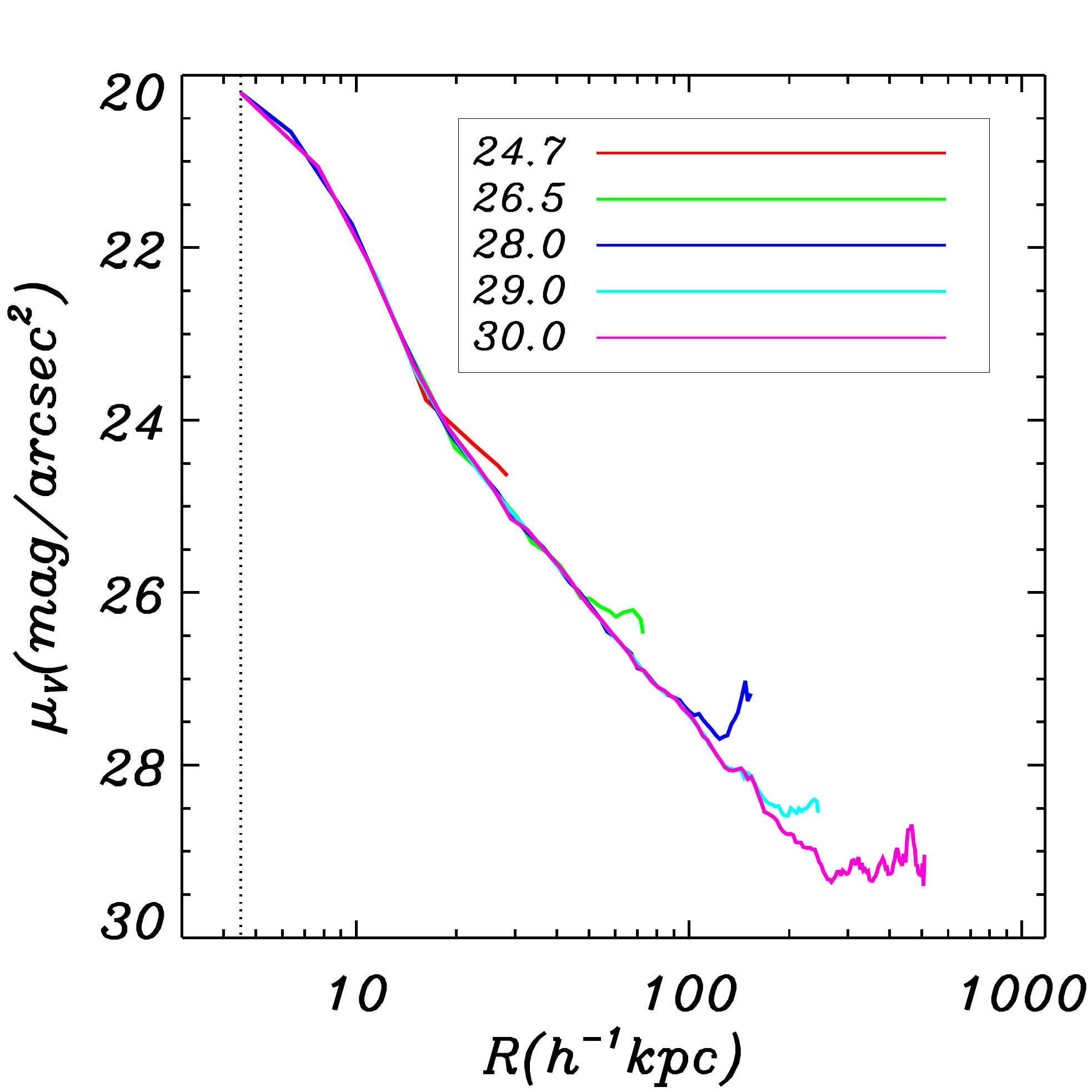}
    \caption{This figure shows the surface brightness profile of the Brightest Galaxy in Simulation at $z\sim0$ with five different SBLs, $\mu_{V,limit}=24.7,\ 26.5,\ 28,\ 29,\ 30 \magarcsec$.
    Vertical line represents the Plummer softening length in the simulation.
    }
    \label{figure_Luminosity_profile}
\end{figure}
%----------------%
%--table_para_SBL--%
\tabletypesize{\footnotesize} \tabcolsep=0.05cm
\begin{deluxetable}{lcccccccc} \tablecolumns{10}
\tablewidth{0pt}
\tablecaption{ Properties of the brightest galaxy in simulation for different surface brightness limits and reconstruction}
\tablehead{\colhead{$\mu_{V,limit}$} & \colhead{$M_*$} & \colhead{$L$} & \colhead{Met} & \colhead{Age(Gyr)} & \colhead{$^{0.1}(g-r)$} & \colhead{Radius} 	}
\startdata
$\magarcsec$ & $10^{12} h^{-1}M_{\odot}$ & $10^{12} h^{-1}L_{\odot}$ & $Log(Z/Z_{\odot})$ & Gyr & magitude & $h^{-1}kpc$  
\\  \ \ \ \ \ \ 24.7$^{(a)}$  & 4.80 & 1.16 & -1.81 & 9.89 & 0.80 & 30.63     
\\  \ \ \ \ \ \ 26.5$^{(a)}$  & 6.15 & 1.40 & -1.91 & 9.95 & 0.81 & 73.80     
\\  \ \ \ \ \ \ 28.0$^{(a)}$ &  7.63 & 1.70 & -2.05 & 9.96 & 0.81 & 154.90   
\\  \ \ \ \ \ \ 29.0$^{(a)}$ &  9.03 & 2.01 & -2.14 & 9.75 & 0.80 & 245.19   
\\  \ \ \ \ \ \ 30.0$^{(a)}$ &  11.9 & 2.67 & -2.26 & 8.53 & 0.79 & 510.73  
\\  \ \ \ \ \ \ 26.65$^{(b)}$  & 6.32 & 1.44 & -1.93 & 9.95 & 0.81 & 81.94        
\enddata
\tablecomments{  The brightest galaxy in simulation is located in the most massive dark matter halo with viral mass $M_{vir}=6.23\times10^{14}h^{-1}M_{\odot}$. 
For galaxy obtained by the reconstruction procedure, the $\mu_{V,limit}=26.65$ here is the surface brightness at the galaxy edge.
This is different from others. 
(a): Brightest galaxy defined by the different surface brightness limits.
(b): Brightest galaxy obtained by the reconstruction procedure with multiple SBLs.
}
\label{tab:para_SBL}
\end{deluxetable}
%------------------%
%--Figure_illustration--%
\begin{figure*}
  \centering  
   \includegraphics[width=0.7\textwidth]{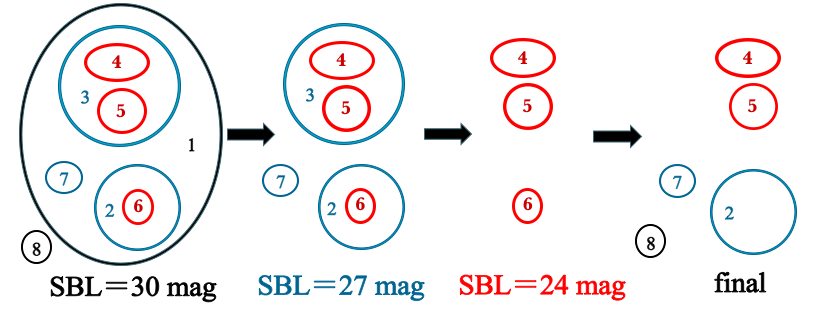}
    \caption{An example for the reconstruction procedure.
    We choose three SBLs, $30 (black),\ 27(blue),\ 24 (red) \magarcsec$.
    The left panel shows the surface brightness profile for the chosen SBLs. 
    It is found that there is only two distinct galaxies, denoted as 1 and 8, for the faintest SBL, $30 \magarcsec$. 
    With a brighter SBL, $27 \magarcsec$, the distinct galaxies are denoted as 2, 3 and 7. 
    The galaxy 8 is so faint that it disappears.
    The galaxy 3 is separated into two galaxies, denoted as 4 and 5, the galaxy 7 is so faint that it disappears, and the galaxy 2 is altered to galaxy 6, with $\rm{SBL}=24 \magarcsec$.
    Rather than the three galaxies (denoted as 4, 5, and 6) defined by the brightest SBL of $24 \magarcsec$, we select galaxies denoted as 2, 4, 5,  7 and 8 for the final distinct galaxies by using our reconstruction procedure in this example.
    }
    \label{figure_illustration}
\end{figure*}
%---------------%
%--Figure_multipe--%
\begin{figure*}
  \centering
  \includegraphics[width=0.27\textwidth, height=0.2\textheight]{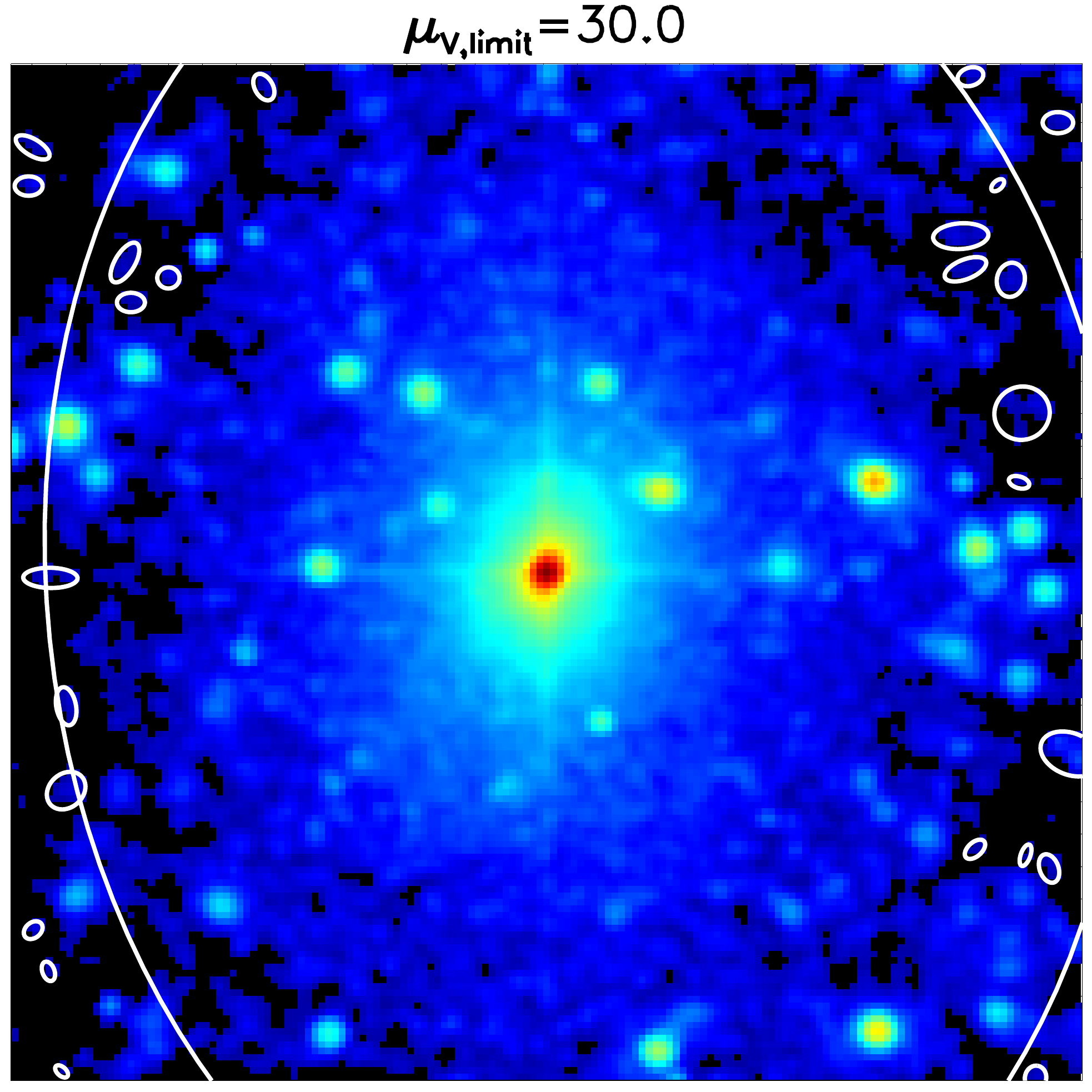}
  \includegraphics[width=0.27\textwidth, height=0.2\textheight]{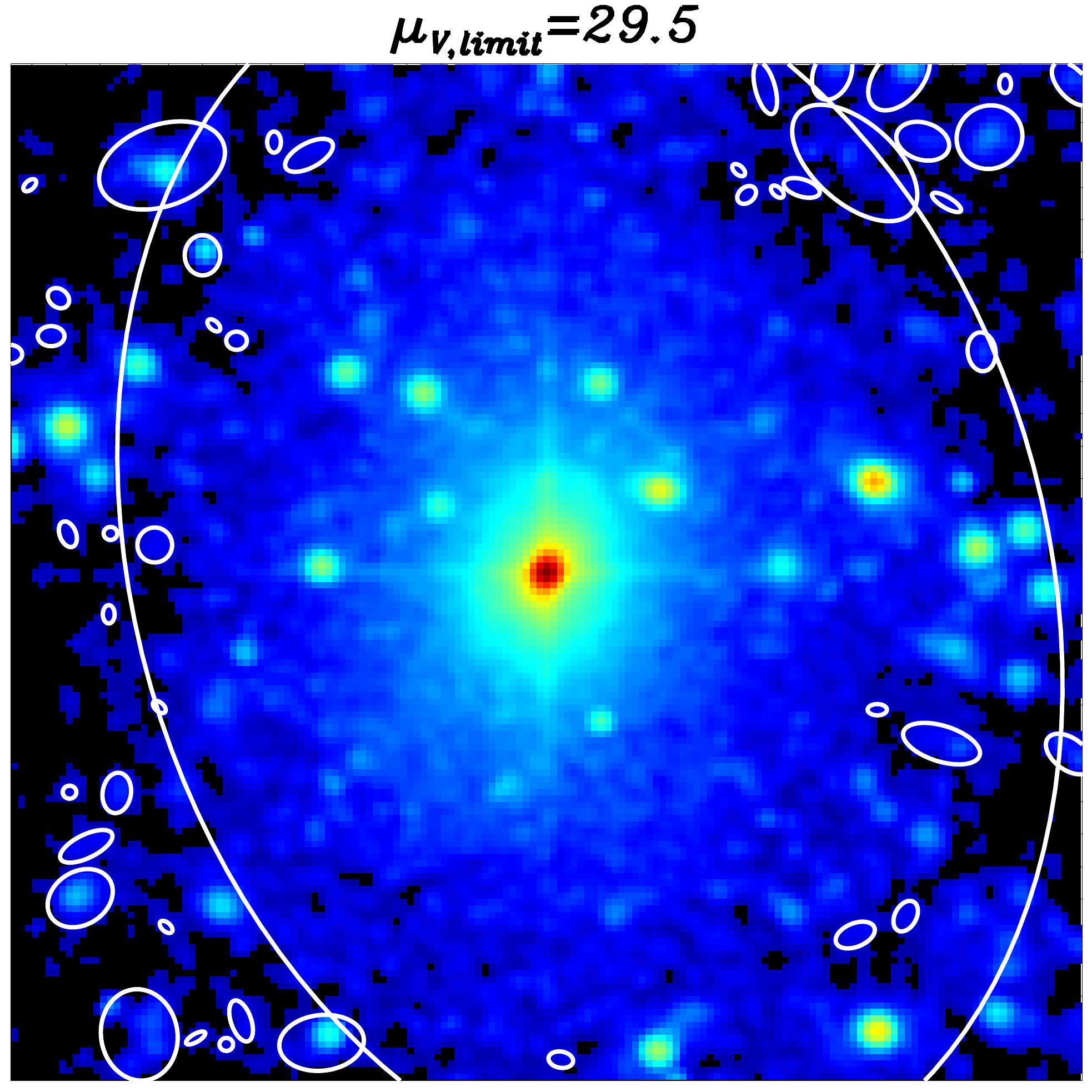}
  \includegraphics[width=0.27\textwidth, height=0.2\textheight]{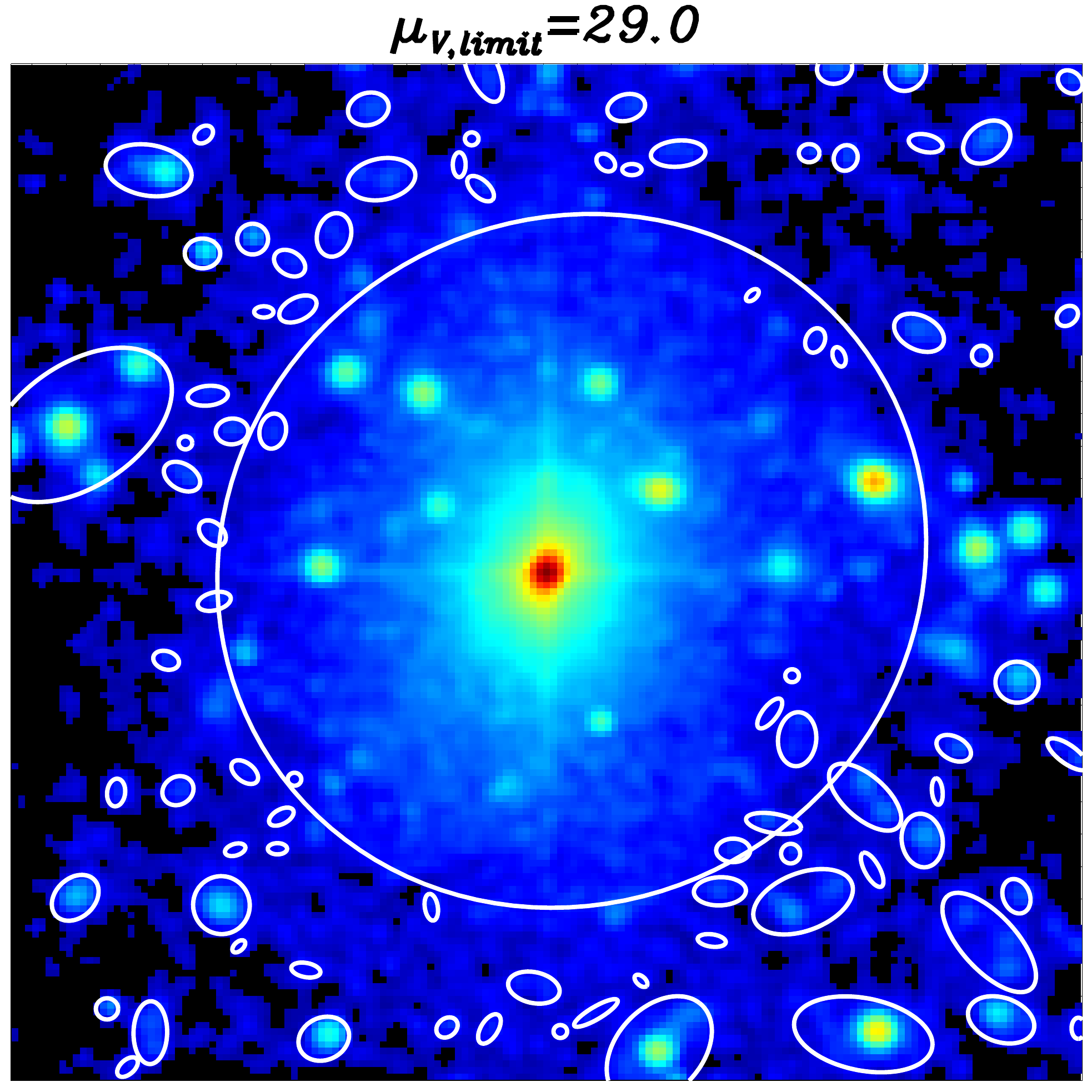}       
  \includegraphics[width=0.27\textwidth, height=0.2\textheight]{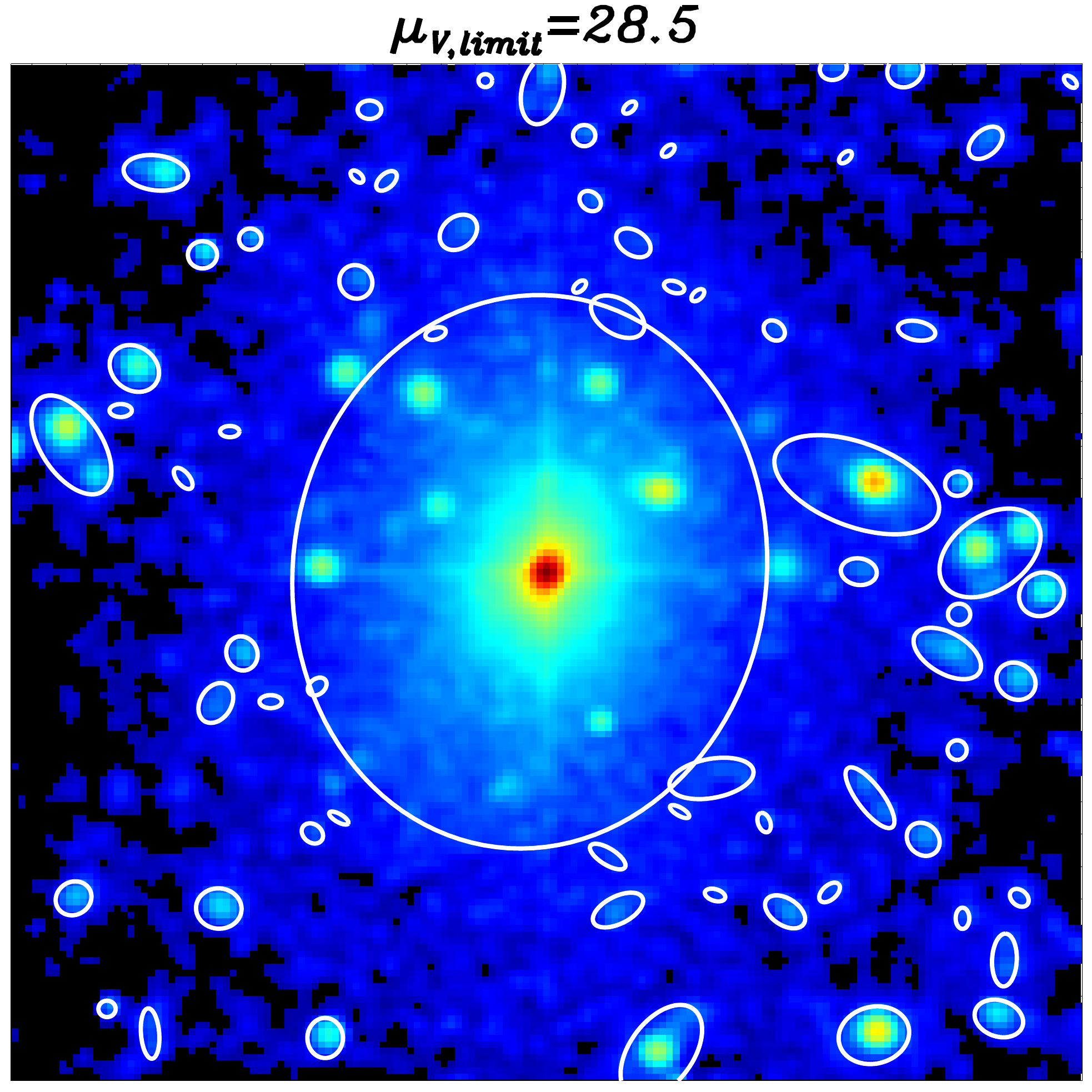}
  \includegraphics[width=0.27\textwidth, height=0.2\textheight]{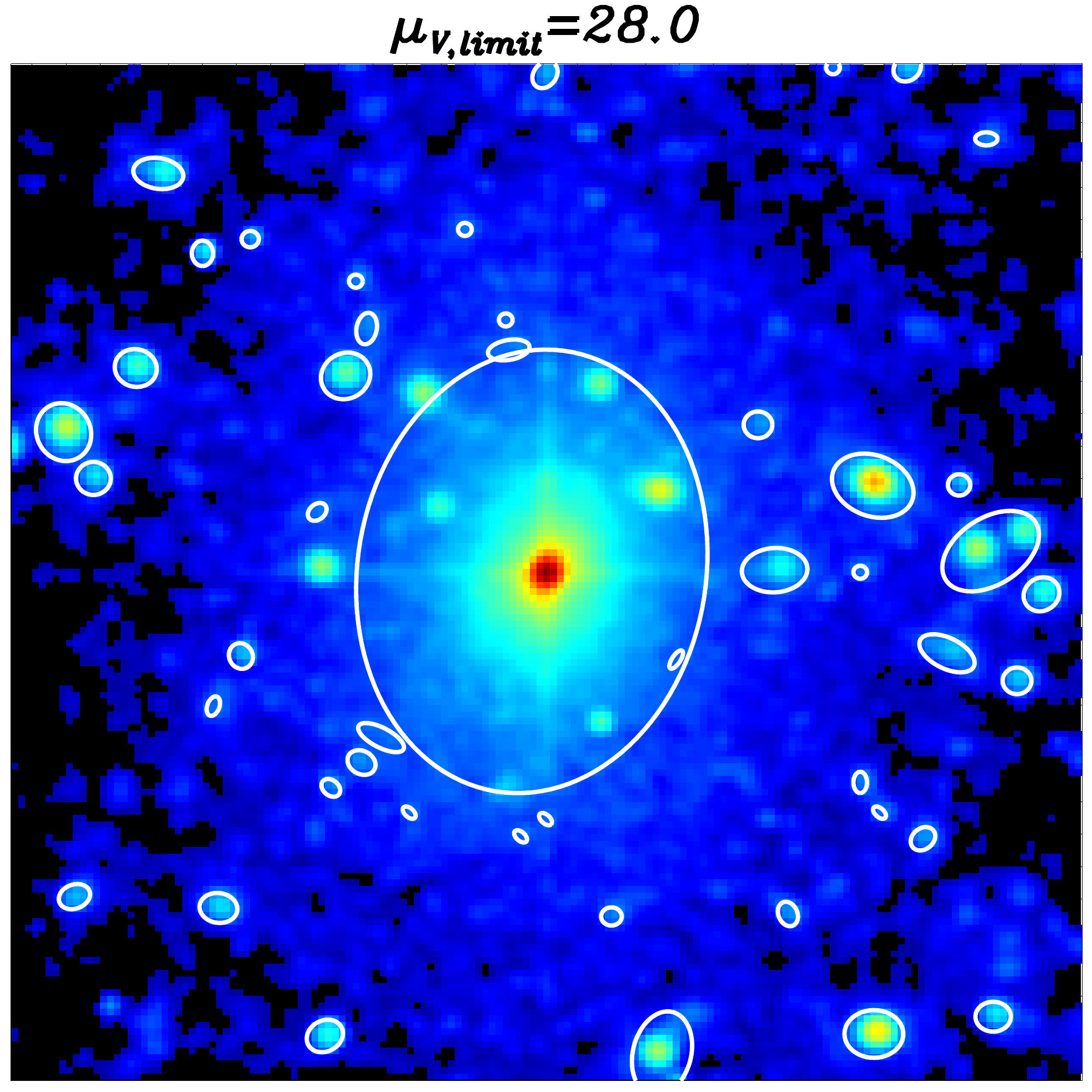}
  \includegraphics[width=0.27\textwidth, height=0.2\textheight]{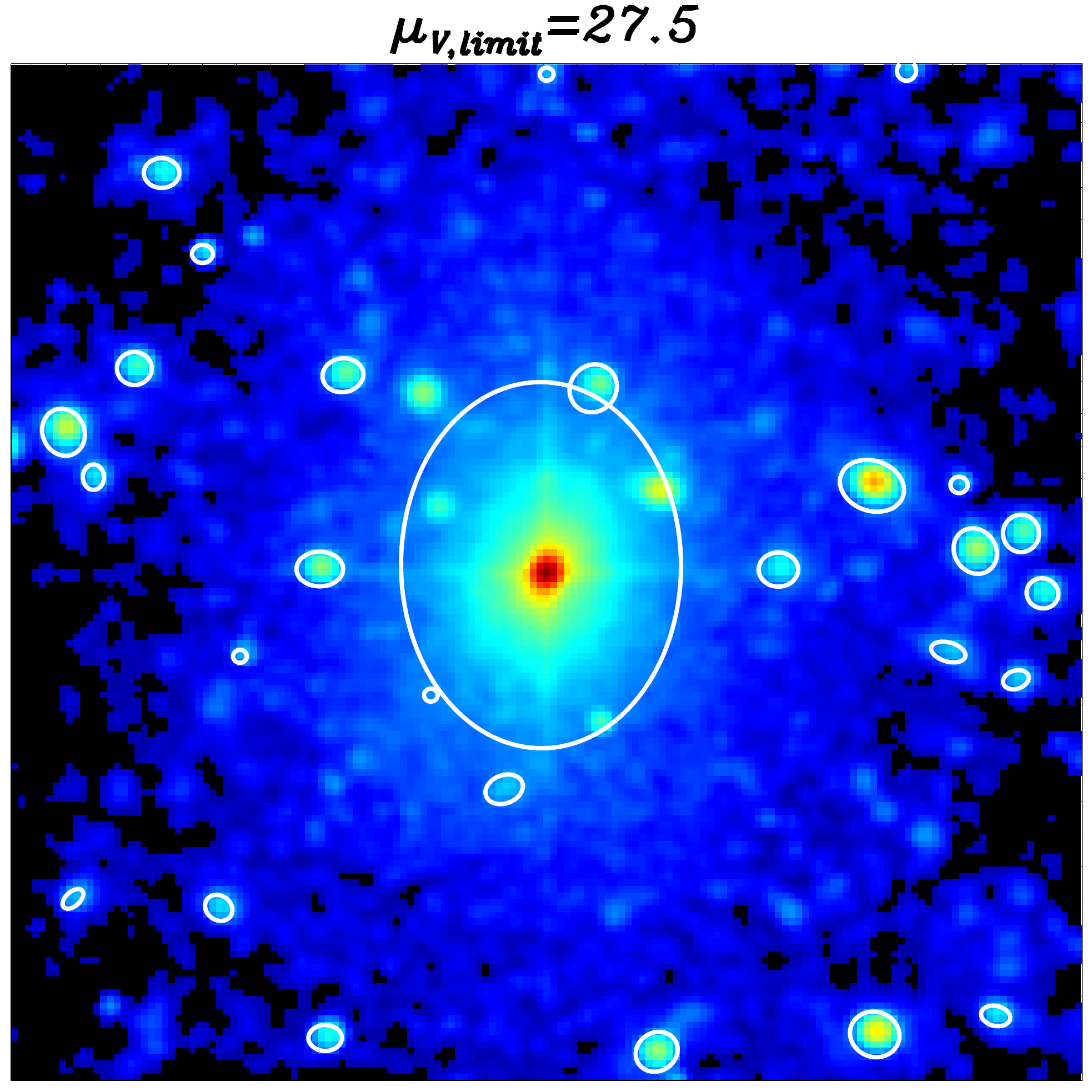}
  \includegraphics[width=0.27\textwidth, height=0.2\textheight]{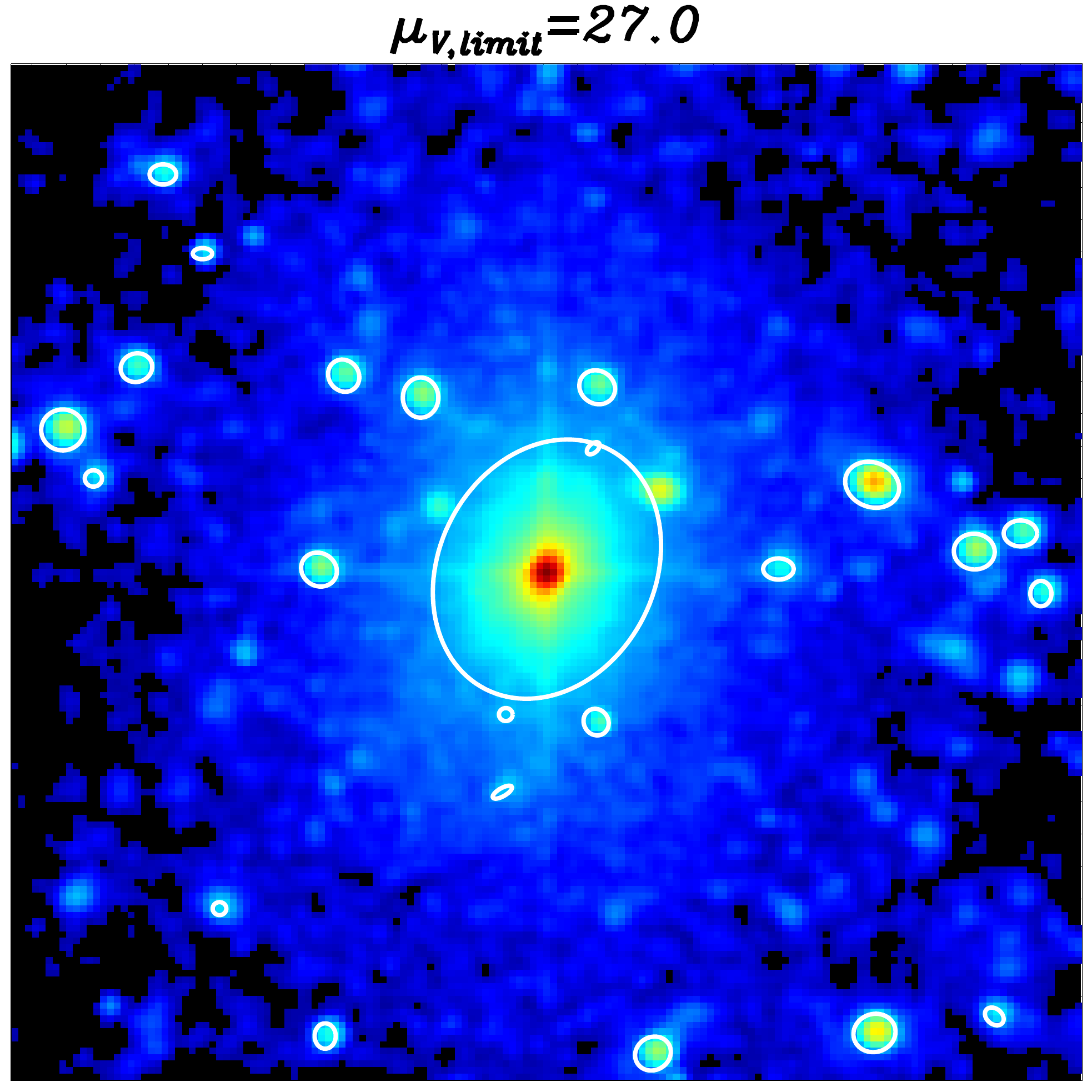}
  \includegraphics[width=0.27\textwidth, height=0.2\textheight]{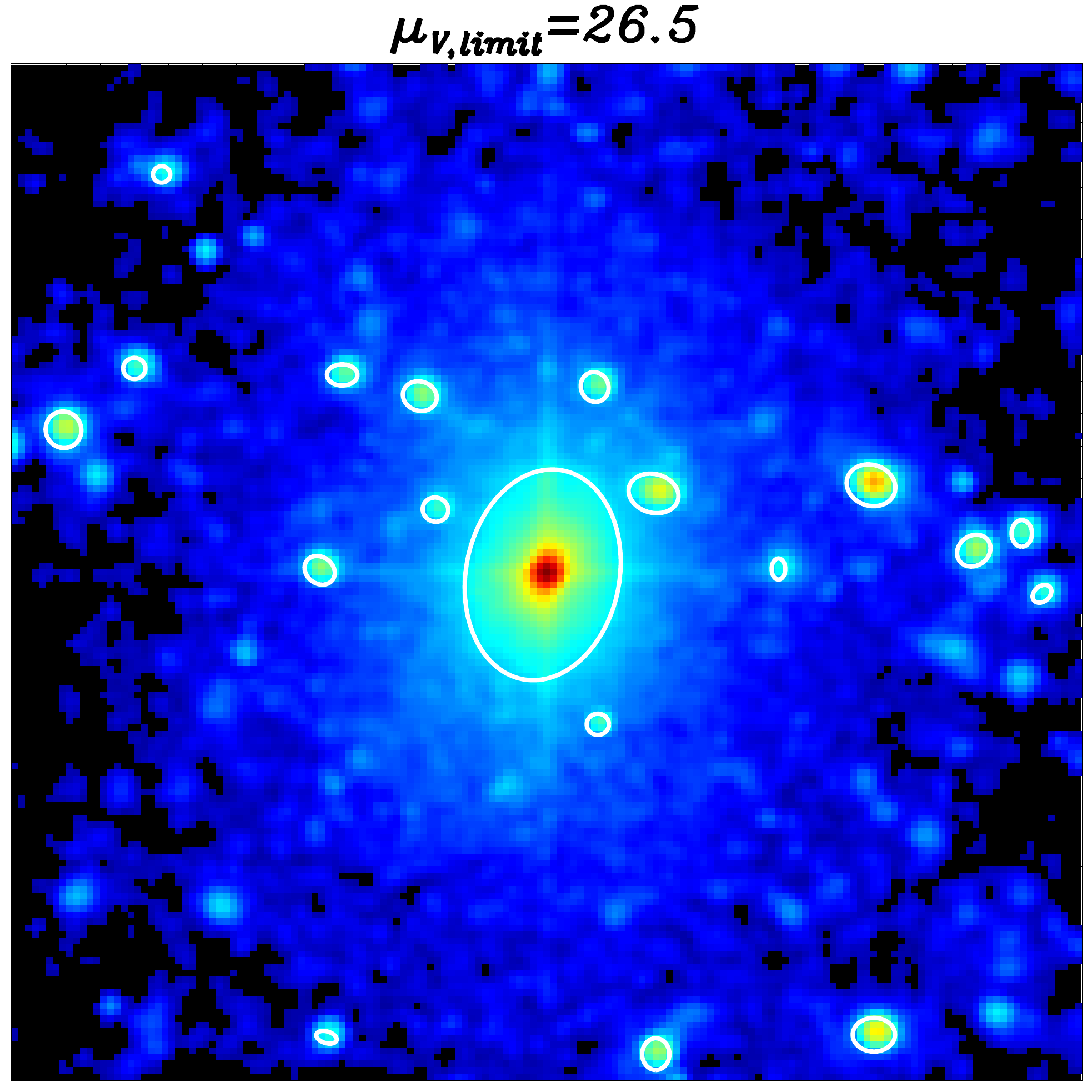}
  \includegraphics[width=0.27\textwidth, height=0.2\textheight]{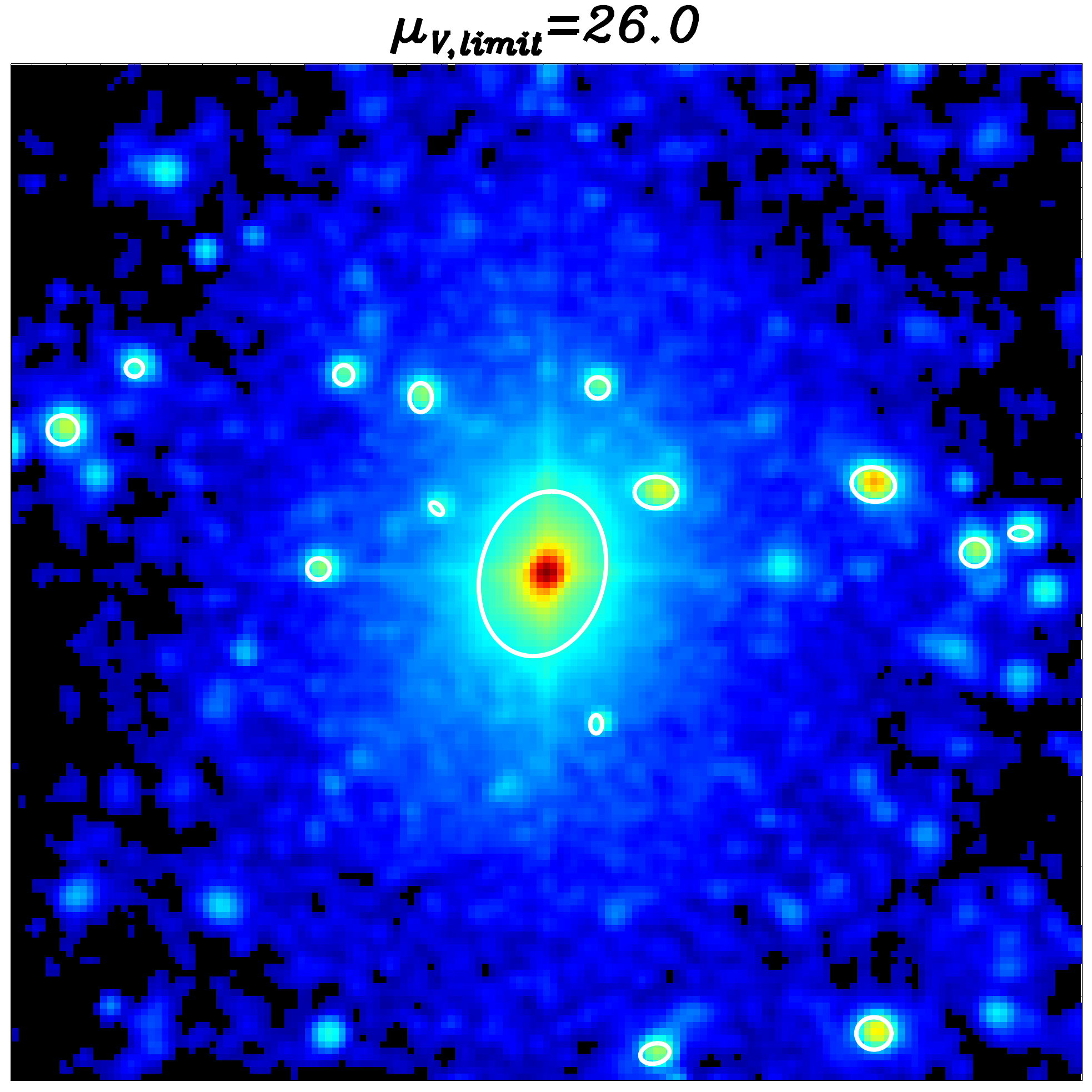}
  \includegraphics[width=0.27\textwidth, height=0.2\textheight]{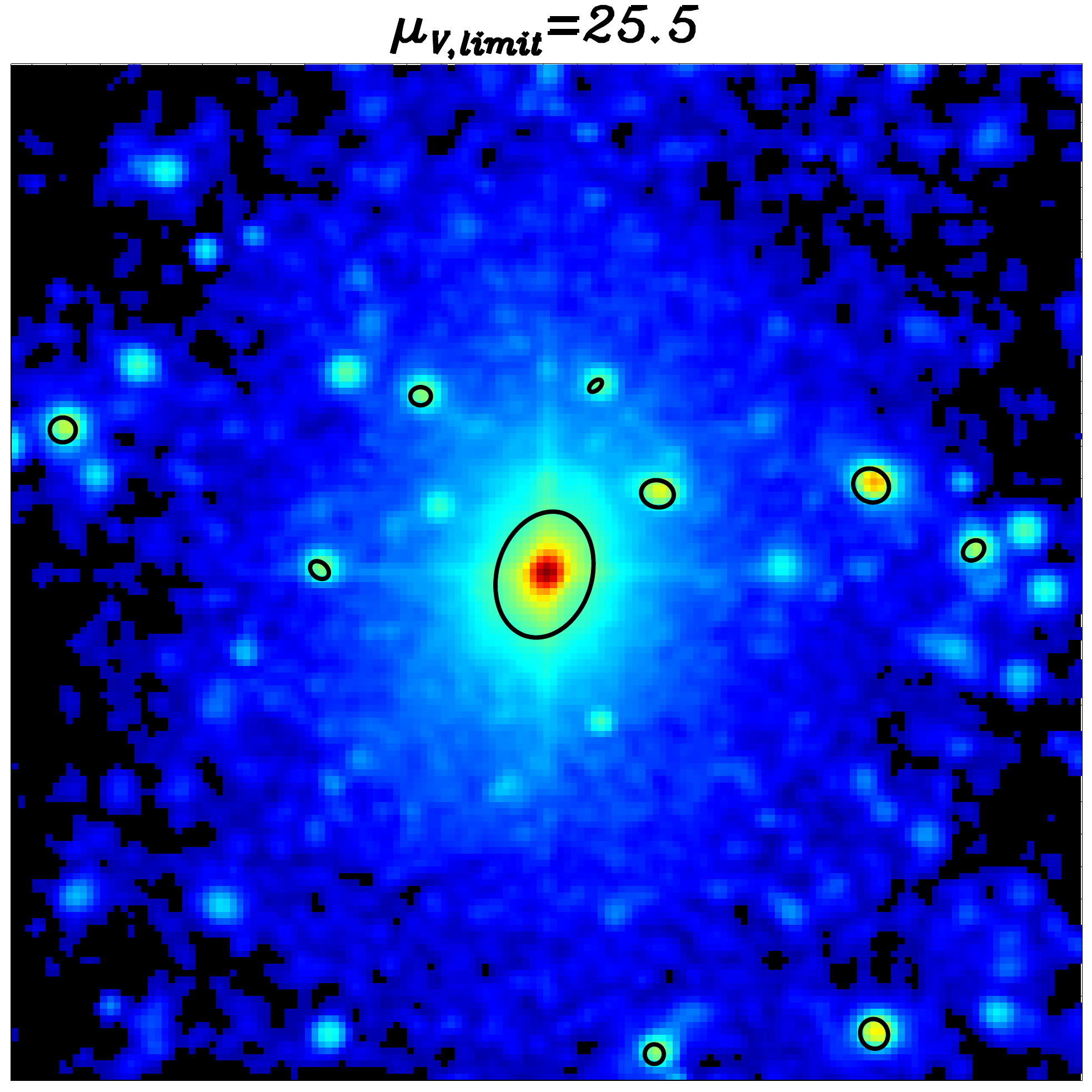}
  \includegraphics[width=0.27\textwidth, height=0.2\textheight]{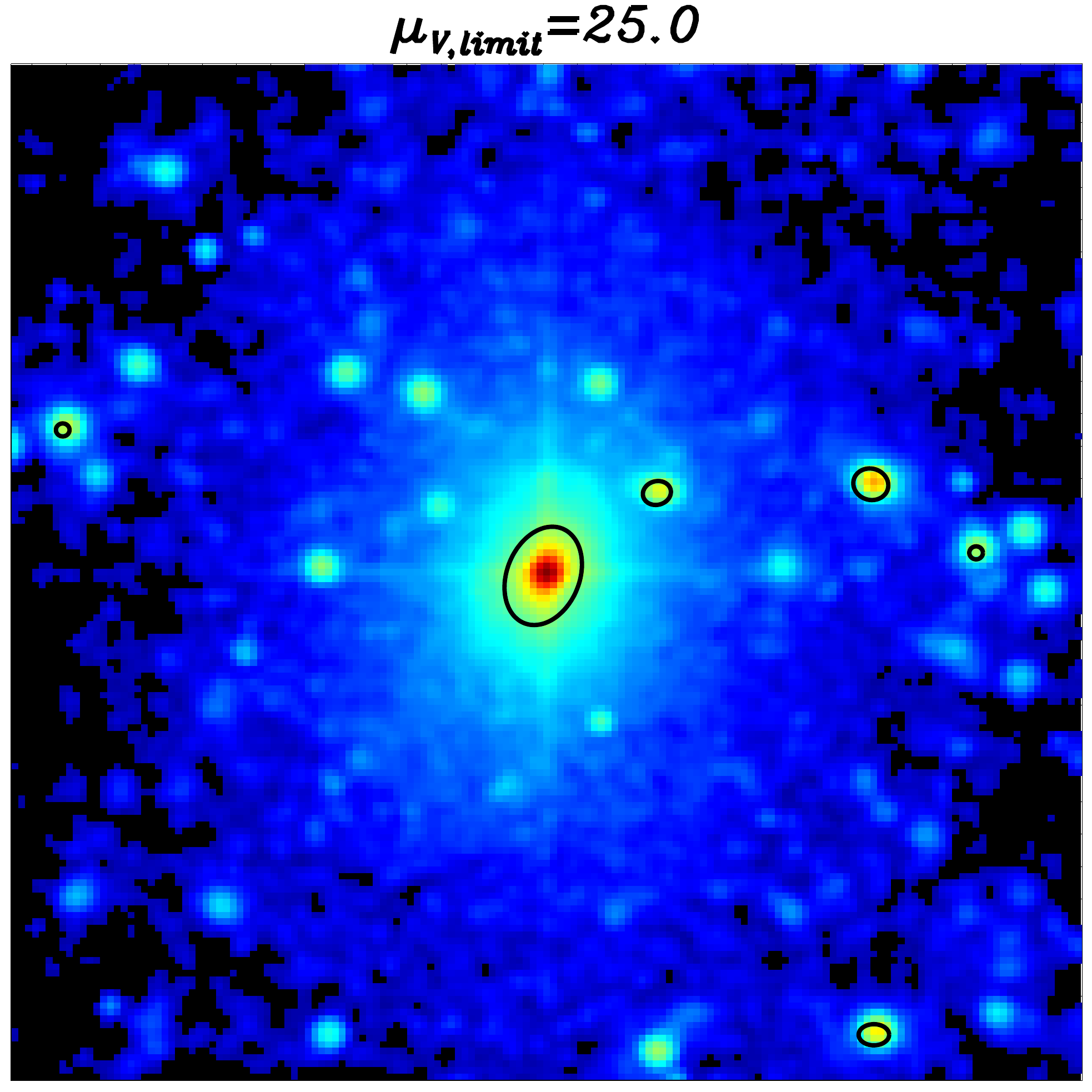}
  \includegraphics[width=0.27\textwidth, height=0.2\textheight]{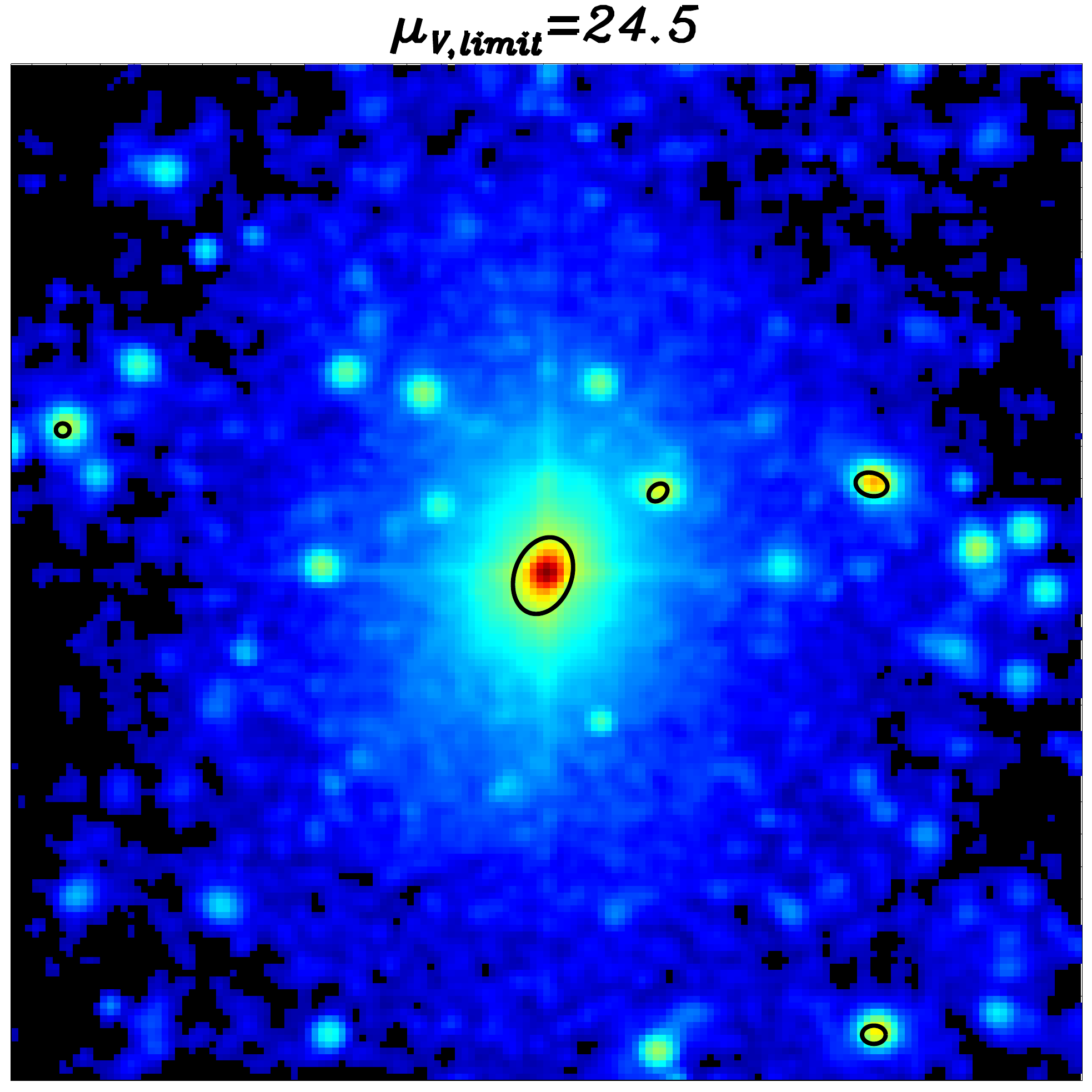}
  \includegraphics[width=0.27\textwidth, height=0.2\textheight]{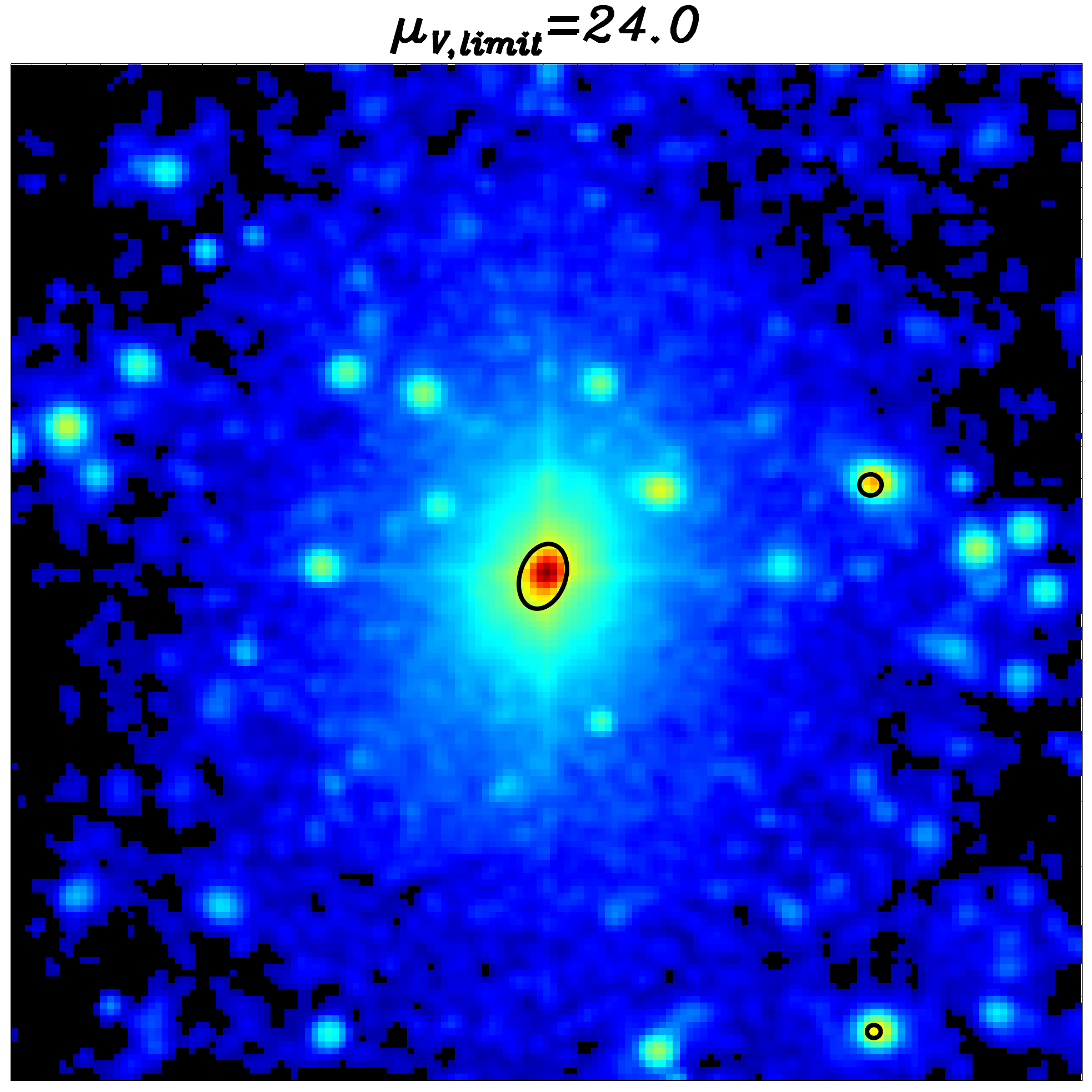}
  \includegraphics[width=0.27\textwidth, height=0.2\textheight]{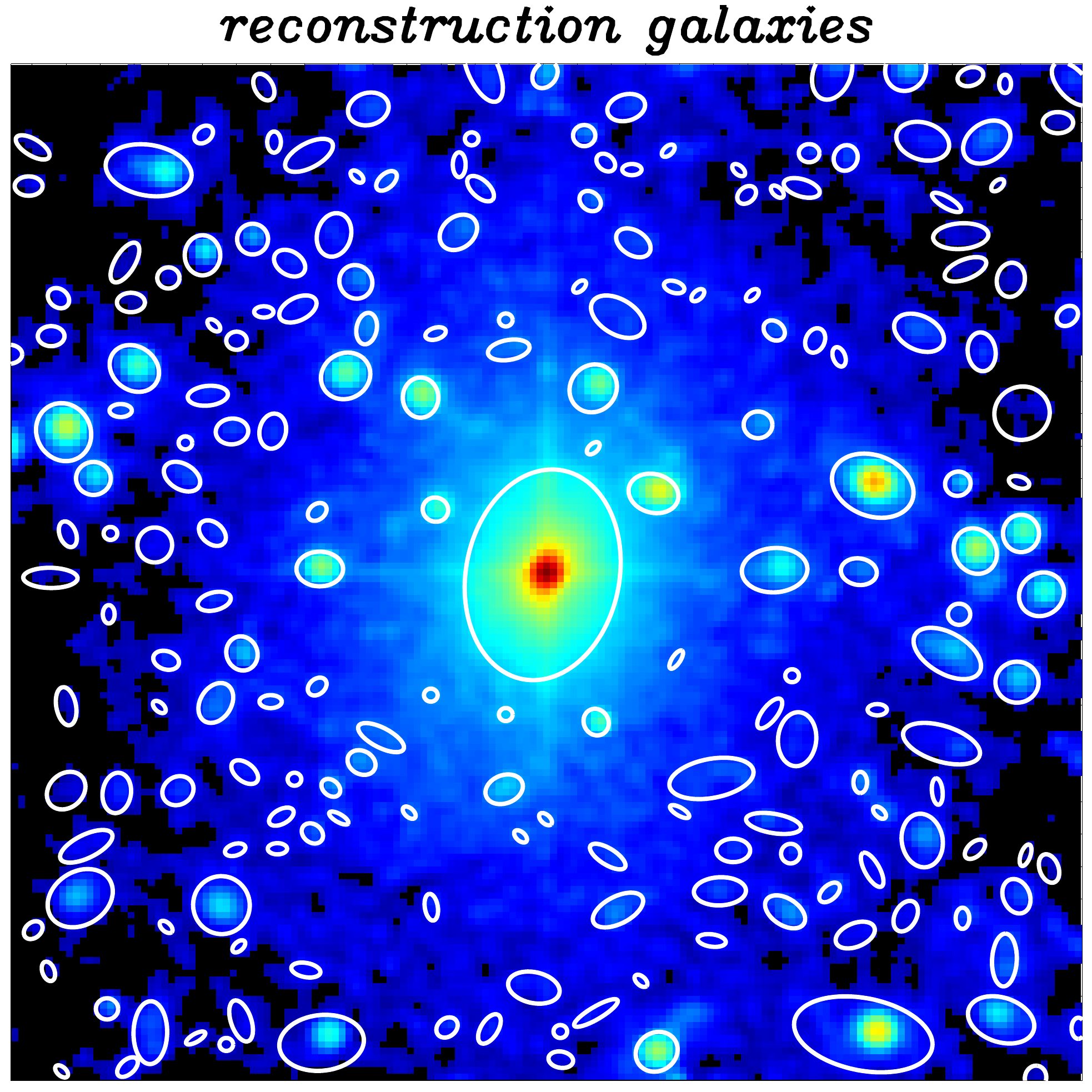}
  \includegraphics[width=0.27\textwidth, height=0.2\textheight]{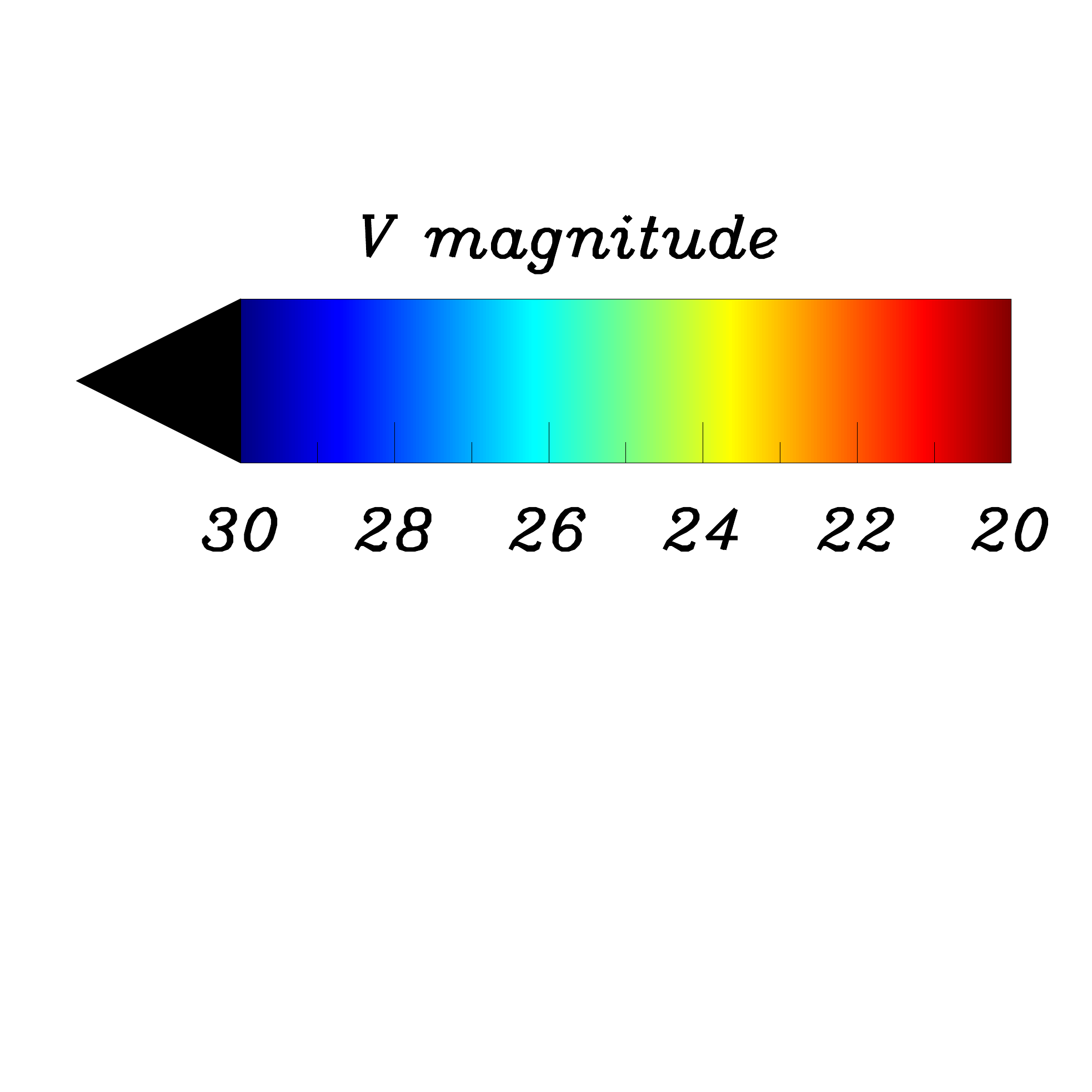}
    \caption{ The galaxies defined by  a series of SBLs with SBL width equal with 0.5 are illustrated. Applying the reconstruction procedure, the reconstructed galaxies are illustrated in the right bottom panel. 
    The white (or black) solid lines are ellipse fitting shape of galaxies.
    The unit of horizontal and vertical axis is $4.5\kpch$.
    Those galaxies are hosted in the most massive dark matter halo.
    We only plot the central region with $700\kpch\times700\kpch$.
    }
    \label{figure_image}
\end{figure*}
%-----------------------------------%
%--Figure_Luminosity_profile_reconstruction--%
\begin{figure}
  \centering  
   \includegraphics[width=0.45\textwidth]{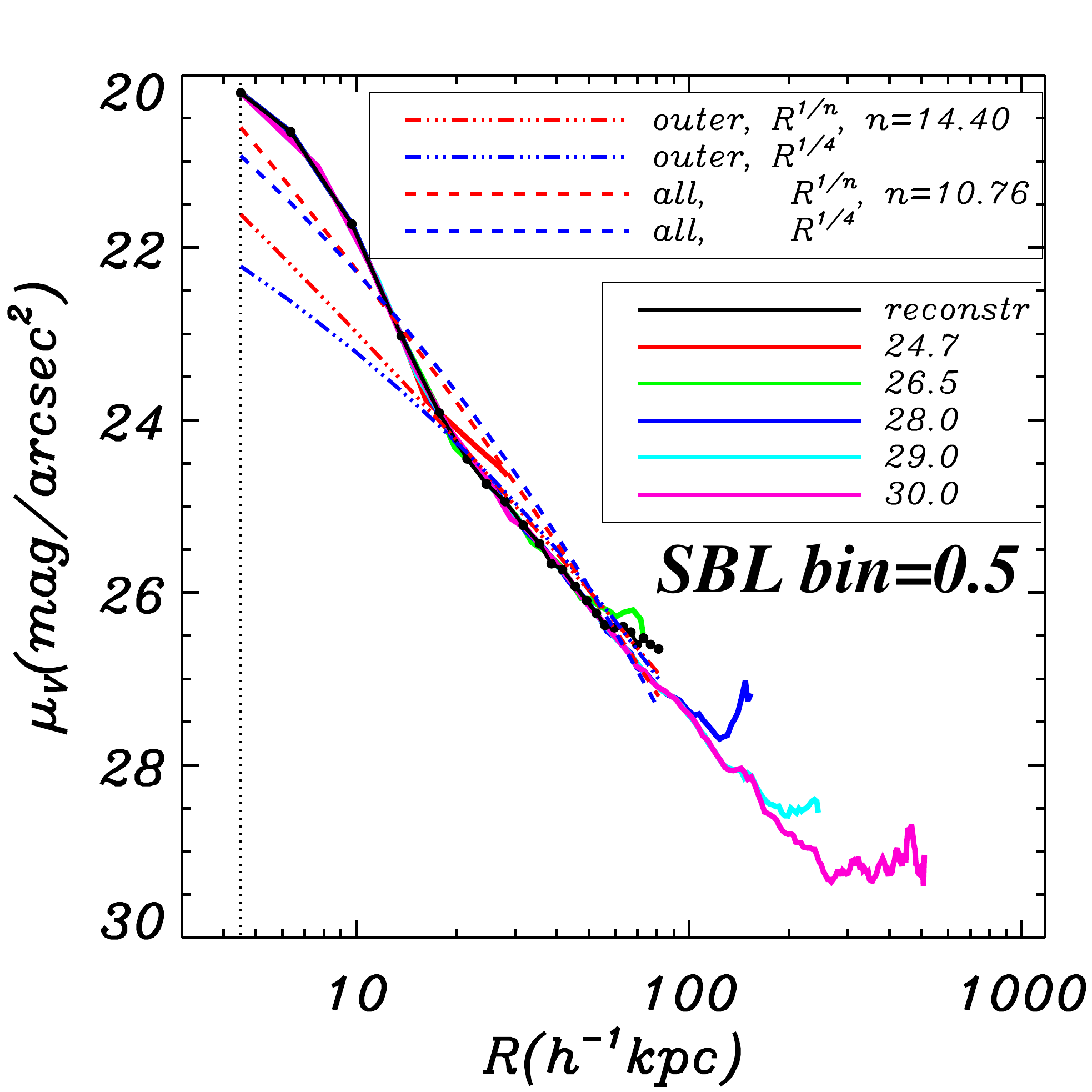}
    \caption{The reconstructed surface brightness profile with SBL bin width equal 0.5 for the same galaxy in Figure~\ref{figure_Luminosity_profile} (black solid line).
    We do the $R^{1/4}$ or $R^{1/n}$ fitting as shown in the legend.
     For comparison with the galaxy in Figure~\ref{figure_Luminosity_profile}, we plot the surface brightness profiles with four different SBLs as shown in Figure~\ref{figure_Luminosity_profile} (solid color lines).
    Vertical line represents the Plummer softening length in the simulation.
    }
    \label{figure_Luminosity_profile_reconstruction}
\end{figure}
%----------------------------------%
%----------------------------------%
Although lacking AGN feedback causes over-cooling problem, producing bluer central galaxies as shown in D14, our method looks realistic and the discrepancy of fraction and evolution of intra-cluster light (ICL) between observations and predictions could be partly reduced \cite[]{Tang2018}.
Stellar magnitude is calculated using the Simple Stellar Population model of \cite{Bruzual&Charlot2003} (hereafter, BC03). 
The calculation of surface brightness and projection method has been described in \cite{Tang2018}.
We only introduce the definition of mock galaxy here.
\par
Firstly, given a PSF width $\omega$ and a CCD pixel size $\alpha$, we obtain the projection image of each FoF group in three different planes, $x-y$, $y-z$, $z-x$.
Surface brightness profile brighter than $30 \magarcsec$ in $V$ band is applied.
Then, applying a given magnitude limit (or surface brightness limit, SBL) in $V$ band, we get the luminosity part above SBL as galaxy components.
Finally, we define grids connected together as one galaxy using a scheme similar to FoF group finder. 
Galaxy properties, such as mass, age, color, metallicity, are calculated by the light-weighted method. 
Particularly, the galaxy shape is defined by the ellipse fitting of all galaxy components.
The major axis is computed by the density profile, which is similar to the method by the surface brightness isophote in previous observational studies.
\par
Color ($g-r$) is defined by BC03 magnitude in SDSS $g$ band minus that in SDSS $r$ band.
Considering comparison between different redshifts, we shift ($g-r$) to color at $z=0.1$, $^{0.1}(g-r)$ using $^{0.1}(g-r)=0.7088-1.3197\lbrack(g-r)-0.6102\rbrack$ \cite[]{Blanton&Roweis2007}.
\cite{Yang2006} simply adopted $^{0.1}(g-r) = 0.83$ to divide galaxies into red and blue subsample, without taking into account the dependence of color on galaxy stellar mass or magnitude.
This method is too simple and results in some red galaxies being assigned as blue galaxies.
As we discussed in Section~\ref{subsec:SRD}, red and blue galaxies show a different spatial distribution.
In fact, it is more reasonable that the division between red and blue galaxies is stellar-mass-dependent \cite[e.g.,][]{vandenBosch2008}.
Therefore, We divide the galaxies into reds and blues using the fitting line of \cite{vandenBosch2008}.
\par
We test the impacts of $\omega$ and $\alpha$ in galaxy definition, and find that smaller $\omega$ and pixel size $\alpha$ make the projection image more similar to the original image of groups. 
Considering the spatial resolution in our simulation, we chose the Plummer softening length as the physical distance of $\omega$ and $\alpha$ for each redshifts.
In addition, as statement in \cite{Tang2018}, the PSF and pixel size effects are more significant for diffuse or under-dense regions.
By applying the selection criteria in next section, the diffuse stellar components should have been removed to ensure clear galaxy boundary.
\par
It is found that galaxy number and physical properties vary significantly in the projected image with different SBLs. 
As shown in Figure~\ref{figure_Luminosity_profile}, we simply explore the surface brightness profile of the brightest galaxy at $z\sim0$ with five different SBLs, [24.7, 26.5, 28, 29, 30]$\magarcsec$ at V band. 
Table~\ref{tab:para_SBL} simply shows the physical properties of the brightest central galaxy defined by different SBLs.
Figure~\ref{figure_Luminosity_profile} shows that the surface brightness profiles of the brightest central galaxy are obviously varied with SBLs. 
It is found that the profiles have fluctuations at large galacto-centric radii for the faintest SBLs, due to the contamination from satellites.
And the profile with the brightest SBL (24.7$\magarcsec$) shows a too small radius, which only includes the central part.
To avoid these problems, we define galaxies by a reconstruction procedure with multiple SBLs (e.g., $24-30 \magarcsec$). 
The final galaxy sample is obtained by combining the results from different SBLs. 
\par
The reconstruction procedure includes following steps.
\begin{enumerate}
\item We define the mock galaxies by a series of SBLs (e.g., [30, 27, 24]$\magarcsec$ with bin width of 3).
\item We compare the mock galaxies defined by two SBLs, for example, [30, 27]$\magarcsec$.
          Those galaxies defined by the faint SBL ($30 \magarcsec$), but not shown in the sample defined by the bright SBL ($27 \magarcsec$), will be included into a temporary sample of reconstructed galaxies.
\item Mock galaxies simultaneously defined by the two SBLs, will be also included into the temporary sample.
         Each galaxy in this sample will be checked if it can be separated into several galaxies or not with the bright SBL.
         If not, the target galaxy defined by the faint SBL will hold its position in the sample.
         Otherwise, the target galaxy will be excluded from the sample and the separated galaxies defined by the bright SBL will be included into the sample.
\item We use this galaxy sample to make the comparison with the one defined by a brighter SBL (repeating second and third steps), and obtain the final sample of reconstructed galaxies.
\end{enumerate}

\par
An example of reconstruction procedure is illustrated by Figure~\ref{figure_illustration}, in particular to show how it works for the second and third steps.
It is found that different galaxies are defined by three different SBLs.
A complete galaxy catalogue in right panel of Figure~\ref{figure_illustration} is set up by the above reconstruction procedure, comparing with the initial state of only one big distinct galaxy and one faint galaxy shown in left panel of Figure~\ref{figure_illustration} (not for real dark matter halo).
\par
We apply a SBL width equal with 0.5 for a SBL range from 24 to 30 $\magarcsec$ in reconstruction procedure.
Figure~\ref{figure_image} illustrates the galaxies with each SBL and the final reconstructed galaxies in the central region of the most massive dark matter halo.
It should be paid attention that the final reconstructed sample has components with vary small radius or low mass, which should not be trusted as galaxies. 
This will be discussed in next section.
The last row of Table~\ref{tab:para_SBL} simply shows the physical properties of brightest galaxy defined by reconstruction procedure.
The surface brightness profile of the reconstructed brightest galaxy is shown in Figure~\ref{figure_Luminosity_profile_reconstruction}.
It is apparent that the problems (fluctuations and small radius) caused by the usage of single SBL have been avoided, and the surface brightness profile becomes smooth.
In general, $26.5 \magarcsec$ is common applied to distinguish galaxy and ICL in observations \cite[e.g.,][]{Feldmeier2004,Presotto2014}.
And the galaxy (especially central brightest galaxy) has a fine surface brightness profile comparing with de Vaucouleurs ($R^{1/4}$) or Se´rsic ($R^{1/n}$) models by using this SBL \cite[e.g.,][]{Gonzalez2005,Zibetti2008}.
As shown in Figure~\ref{figure_Luminosity_profile}, there is feature of fluctuation on the edge region of the central galaxy.
After the reconstruction, we can reproduce the galaxy size defined by SBL$=26.5 \magarcsec$, and the galaxy has a flat surface brightness profile which looks similar to a cD galaxy, as shown in Figure~\ref{figure_Luminosity_profile_reconstruction}. 
As one can see, the impact by neighbor galaxies has been removed.
The profile in the outer region is well fitted by $R^{1/4}$ or $R^{1/n}$ models.
On the other hand, the bulge is much brighter and its surface brightness profile exceeds the model prediction, causing by over-cooling problem and possible over-merging problem in our simulation. 

%------------------%
%--Figure_selection_1--%
\begin{figure}
  \centering  
   \includegraphics[width=0.45\textwidth]{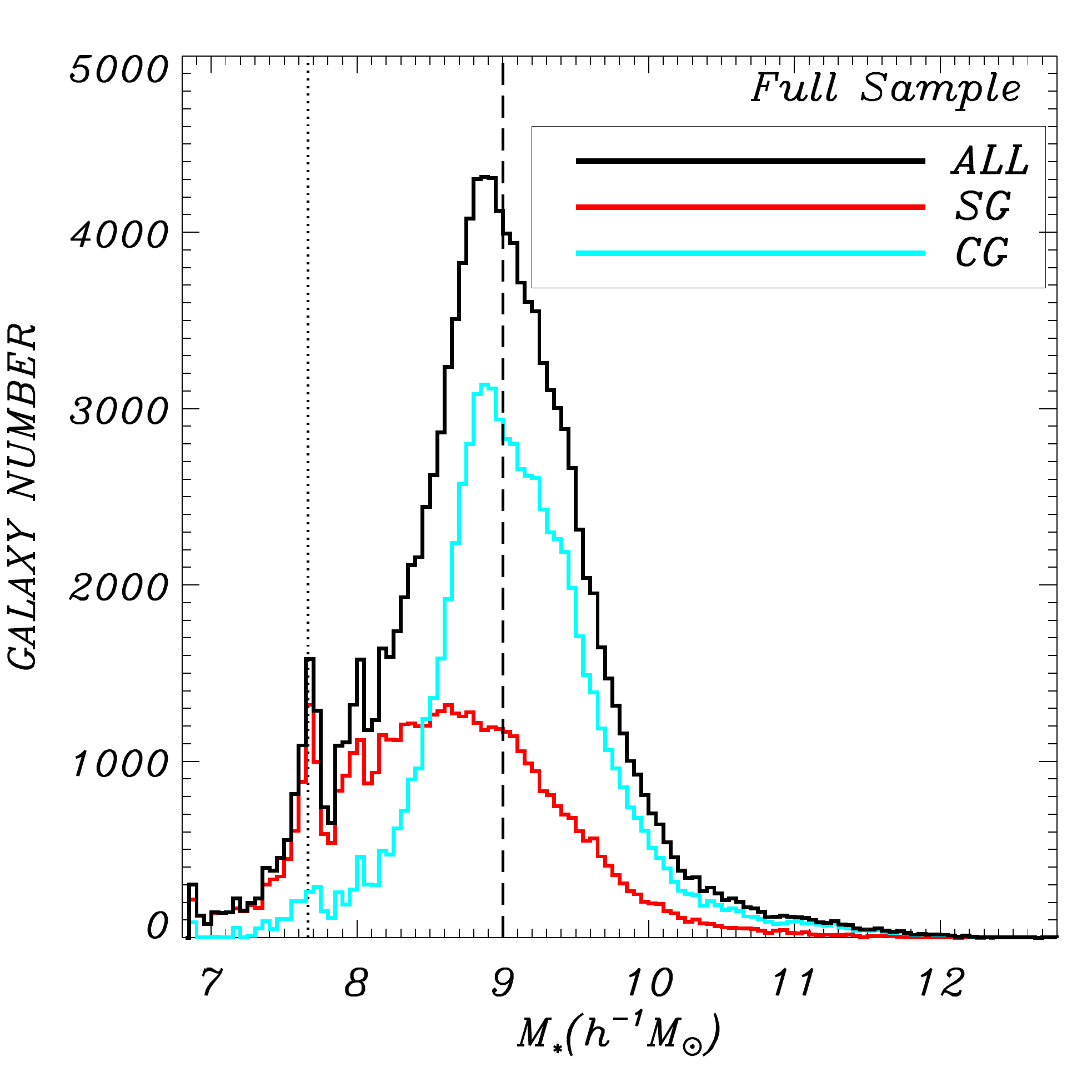}
   \includegraphics[width=0.45\textwidth]{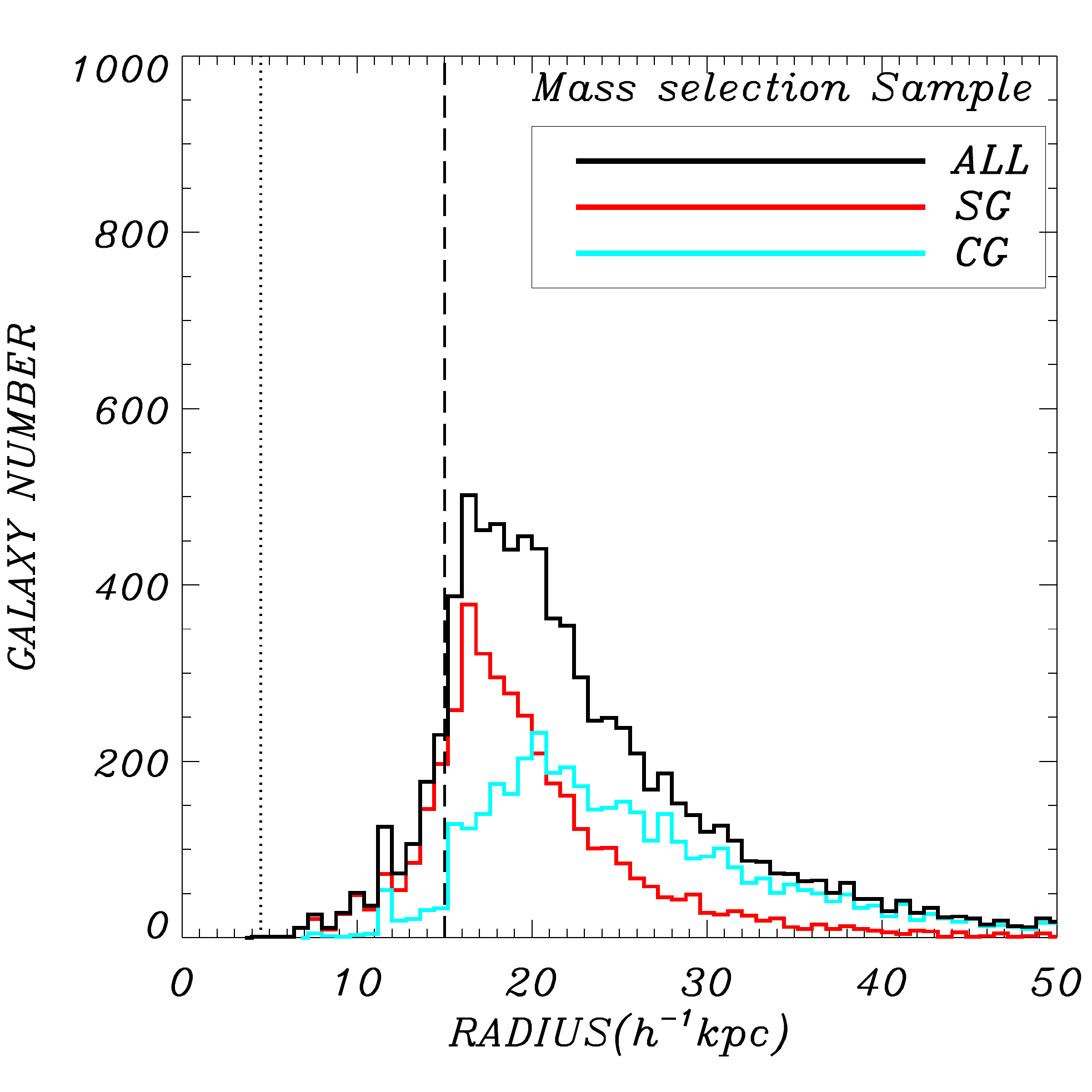} 
    \caption{ Top panel shows the stellar mass function of the different kind of reconstructed galaxy samples with SBL bin width equal 0.5, without any mass or size resolution criteria (i.e., full sample).
    Bottom panel shows radius distribution of the different kind of reconstructed galaxy samples with same SBL bin width (0.5), but including the galaxy mass resolution criterion $M_*>9.2\times10^9 \Msunh$ for centrals and $M_*>4.6\times10^9 \Msunh$ for satellites  (i.e., mass selection sample) as discussion in Section~\ref{sec:selection}.
    The cyan, red and black solid lines mean satellite, central and member (i.e., central+satellite) galaxy samples, respectively.
    The vertical dotted lines in those two panels represent the mass resolution and  Plummer softening length in the simulation, respectively.
    The vertical long dashed lines in those two panels represent $M_*=10^9 \Msunh$ and $Radius=15\kpch$, respectively.
    All the plots are for the results in $x-y$ plane.
    }
    \label{figure_selection_1}
\end{figure}
%-----------------%
%--Figure_selection--%
\begin{figure}
  \centering  
   \includegraphics[width=0.45\textwidth]{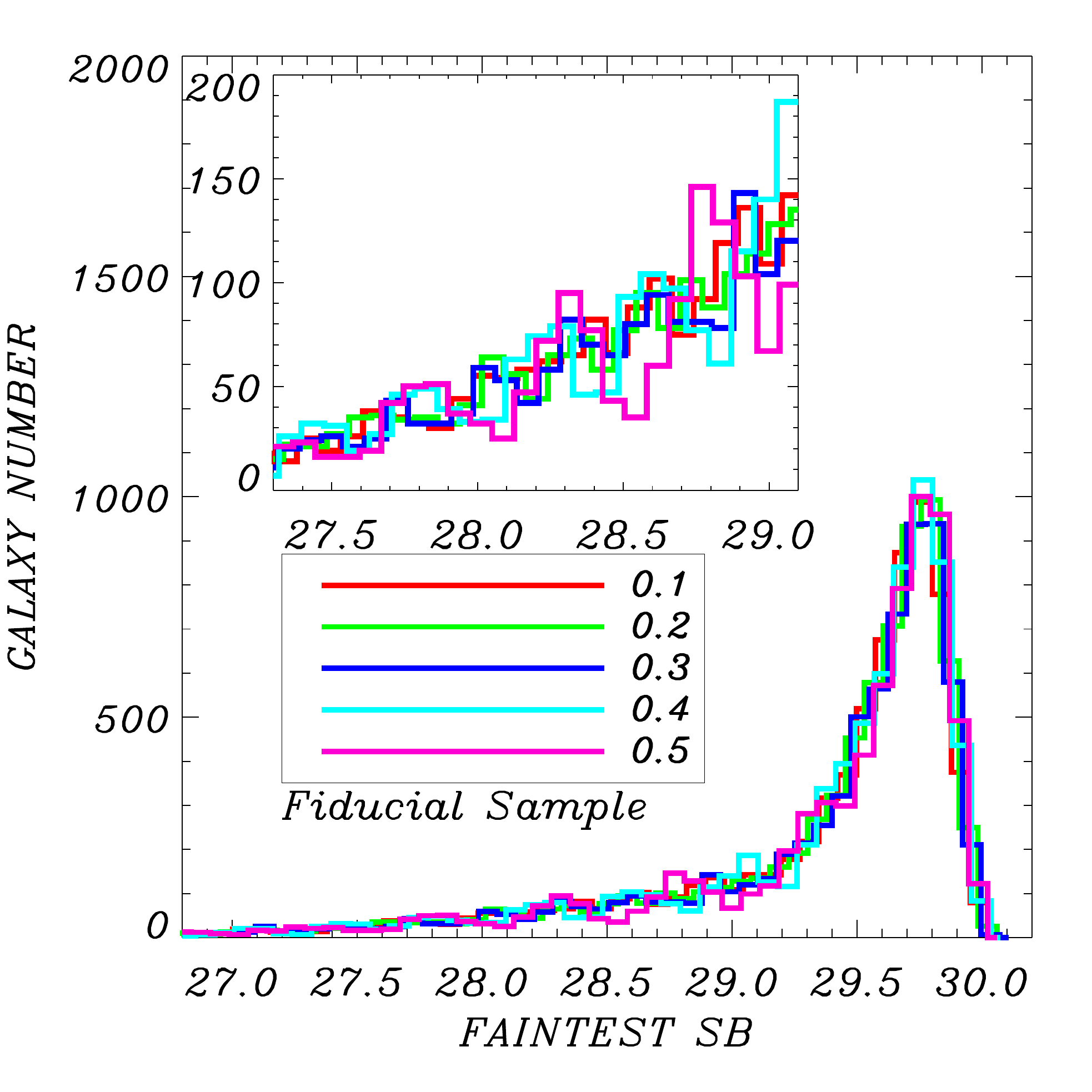}
    \caption{This figure shows the dependence of fiducial galaxy number on the faintest surface brightness (faintest SB), i.e., the lowest value of surface brightness profile of galaxies, which is the mean surface brightness on the edge region of galaxies, with different SBL bin widths, [0.1, 0.2, 0.3, 0.4, 0.5], as shown by the legend.
     %The fiducial sample is restricted by the mass and {\bf size resolution criteria} as discussion in Section~\ref{sec:selection}. 
      The fiducial sample is restricted by the mass and size resolution criteria as discussion in Section~\ref{sec:selection} ( $M_{cen,*}>9.2\times10^9 \Msunh$, $M_{sat,*}>4.6\times10^9 \Msunh$ and $Radius>18 \kpch$).
    All the plots are for the results in $x-y$ plane.
    }
    \label{figure_selection_2}
\end{figure}
%-----------------%

\par
As stated in \cite{Kang2007} and \cite{Wang2014a}, the ``orphan'' galaxies (i.e., galaxies with unresolved or tidally disrupted small subhalos) are important for the satellite alignment signal.
The ``orphan'' galaxies are mostly reside in the inner region of host halos.
Those galaxies are commonly included by the main subhalos defined by the substructure finder.
Using our reconstruction procedure, these galaxies can be clearly distinguished.
It is also worth noting, utilizing state-of-the-art cosmological simulations, it is found that most disruption (or “orphan galaxy”) is numerical in origin\cite[e.g.,][]{vandenBosch2018,vandenBosch&Ogiya2018}. 
We will come back to this problem using present-day simulations, e.g., Illustris simulation.
%-----------------%
%--selection criteria--%
\subsection{the refinement of galaxy sample}
\label{sec:selection}
%------------------%

The Plummer softening length is $\epsilon=4.5 \kpch$, and each stellar particle has a mass of  $4.62\times10^7 \Msunh$ in our simulation.
Commonly,  substructures with stellar mass and size under a given value, associated with resolution in simulation, are not treated as galaxies.
%----Sample Defination------%
\begin{table*}[htbp]
	\centering  
	\caption{ Sample Definitions } 
	\begin{tabular}{|c|c|}  
		\hline  
		 Full Sample & All galaxies in the simulation (top panel of Figure~\ref{figure_selection_1}). \\  
		\hline
		Mass selection Sample & Galaxies with $M_{cen,*}>9.2\times10^9 \Msunh$, $M_{sat,*}>4.6\times10^9 \Msunh$ (bottom panel of Figure~\ref{figure_selection_1}). \\
		\hline
		Fiducial Sample & Galaxies with  $M_{cen,*}>9.2\times10^9 \Msunh$, $M_{sat,*}>4.6\times10^9 \Msunh$ and $Radius>18 \kpch$ (Figure~\ref{figure_selection_2}). \\
		\hline
	       Comparative Sample &  Galaxies with $M_{cen,*}>9.2\times10^9 \Msunh$, $M_{sat,*}>4.6\times10^9 \Msunh$, $Radius>18 \kpch$, $m_r<22.2 \ \rm magnitude$, \\
		& and located in dark matter halos with $M>10^{12}h^{-1}M_{\odot}$ (Figure~\ref{figure_probability_function_all} and \ref{figure_probability_function_type}). \\
		\hline
	\end{tabular}
       \label{tab:define}  
\end{table*}
%----------%
\par
As shown in the top panel of Figure~\ref{figure_selection_1} for the full sample without any selection, the stellar mass function of central galaxies peaks at $M_{\ast}\sim 10^9 \Msunh$, and declines sharply toward the lower mass end.
Furthermore, for satellite galaxies, it shows unreasonable peaks in $M_{\ast}< 10^9 \Msunh$. 
Considering the computational accuracy of centrals' axis and reliability of galaxies \cite[e.g.,][]{Bett2007,Wang2015}, we set a mass resolution criterion to select only central and satellite galaxies more massive than $9.2\times10^9 \Msunh$, and $4.6\times10^9 \Msunh$, respectively, which are approximately 200 and 100 times mass resolution, and enough for resolving the structures of galaxies \citep{Bett2007}. 
Table~\ref{tab:define} shows the sample definitions.
On the other hand, the criteria are above the peak position of mass function, which guarantee the completeness of galaxy sample.
\par
In the bottom panel of Figure~\ref{figure_selection_1}, we plot the galaxy radius number distribution for the sample with the mass resolution criterion applied.
It is found that galaxy radius distribution peaks at $\sim 15 \kpch$. 
With the mass resolution criterion, we set a size criterion to select only galaxy with radii larger than $18 \kpch$, approximately 4 times spatial resolution, which is slightly larger than the peak value of galaxy radius distribution.
\par
The lower limit of radius and particles number for our selected galaxies ($r_h/\epsilon>4$ and $N>200(100)$) is slightly smaller than the criteria in \cite{vandenBosch&Ogiya2018} ($r_h/\epsilon>6.9$, $N>250$). 
But keep in mind that our method is more conservative since it cuts off more particles close to the edges of each galaxy. 
Thus these slightly lower criteria will not bring more fake galaxies. 
\par
The SBL bin width is associated with the observational magnitude resolution. 
Considering two selection criteria above, we reconstruct galaxies with five SBL bin widths, [0.1, 0.2, 0.3, 0.4, 0.5], to investigate the selection effect of SBL bin width.
We present the dependence of galaxy number distributions on the faintest surface brightness in the most outer region of galaxies, i.e., the lowest value of surface brightness profile of galaxies, which is the mean surface brightness on the edge region of galaxies shown in Figure~\ref{figure_selection_2}.
It is found that the distributions are a bit varied with different SBLs. 
It shows small repeated fluctuations with intervals almost equal to bin width. 
The fluctuations are more obvious while the bin width is larger. 
Amplitude of fluctuation is reduced with smaller SBL bin width, meaning that a smaller SBL bin width, brings smoother distribution of faintest surface brightness.
Consequently, reconstruction using a smaller SBL bin width can produce more realistic mock galaxies.
We also check the impact of SBL bin widths on galaxy stellar mass and radius, and find that the distributions are similar for different bin widths.
In the following study, we reconstruct the galaxies with SBL bin width equal with 0.1.

%--Results--%
\section{Results}\label{sec:results}
%---------------------------%
%--Figure_probabiliy_function_all--%
\begin{figure}
  \centering
  \includegraphics[width=0.45\textwidth, height=0.4\textheight]{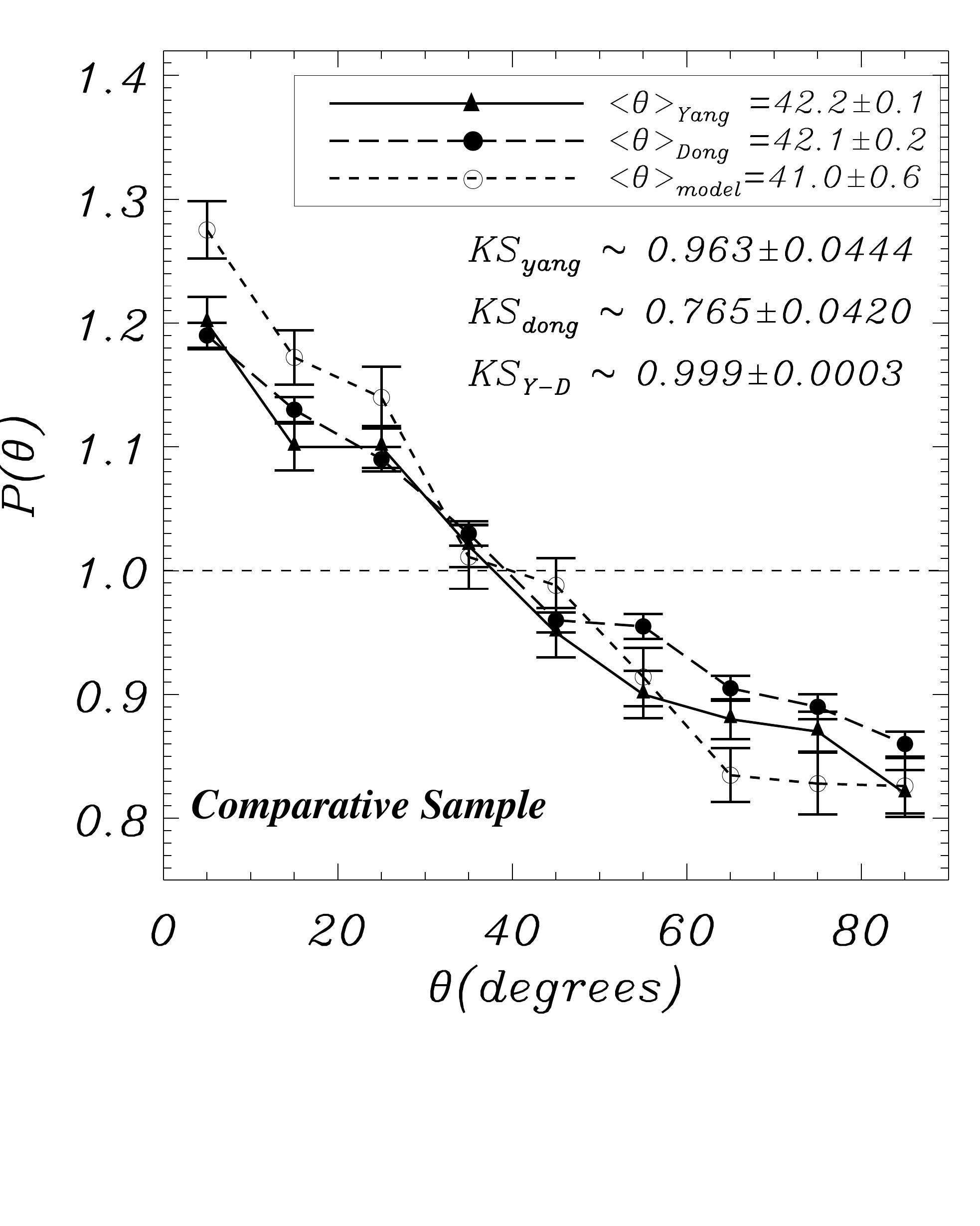}        
  \vspace{-1.5cm}
    \caption{Predicted alignment of satellite galaxies at $z\sim0$ of  comparative galaxy sample (short dashed line with blank circle), observational alignment (solid line with solid triangle) and previous theoretical prediction (long dashed line with solid circle). 
    The average angles $<\theta>$ are shown in the legends.
    We show the standard deviation of the average angles.
   The statistical error is Poisson error.
   The $KS_{Yang}$ and $KS_{Dong}$ are the KS probability of drawing our prediction from the results in Y06 and in D14, respectively.
    The $KS_{Y-D}$ is the KS probability between the results in Y06 and in D14.
    Galaxy sample is combined with those on three projected planes.}
    \label{figure_probability_function_all}
\end{figure}
%----------------------------%
We use the angle in projection between the major axis of central galaxy and the connecting line between center of a satellite and host central to express the satellite spatial distribution. 
Central galaxy is defined as the most massive galaxy in each dark matter halo. 

%-----------------------------%
%--Figure_probabiliy_function_type--%
\begin{figure*}
  \centering
  \includegraphics[width=0.45\textwidth, height=0.4\textheight]{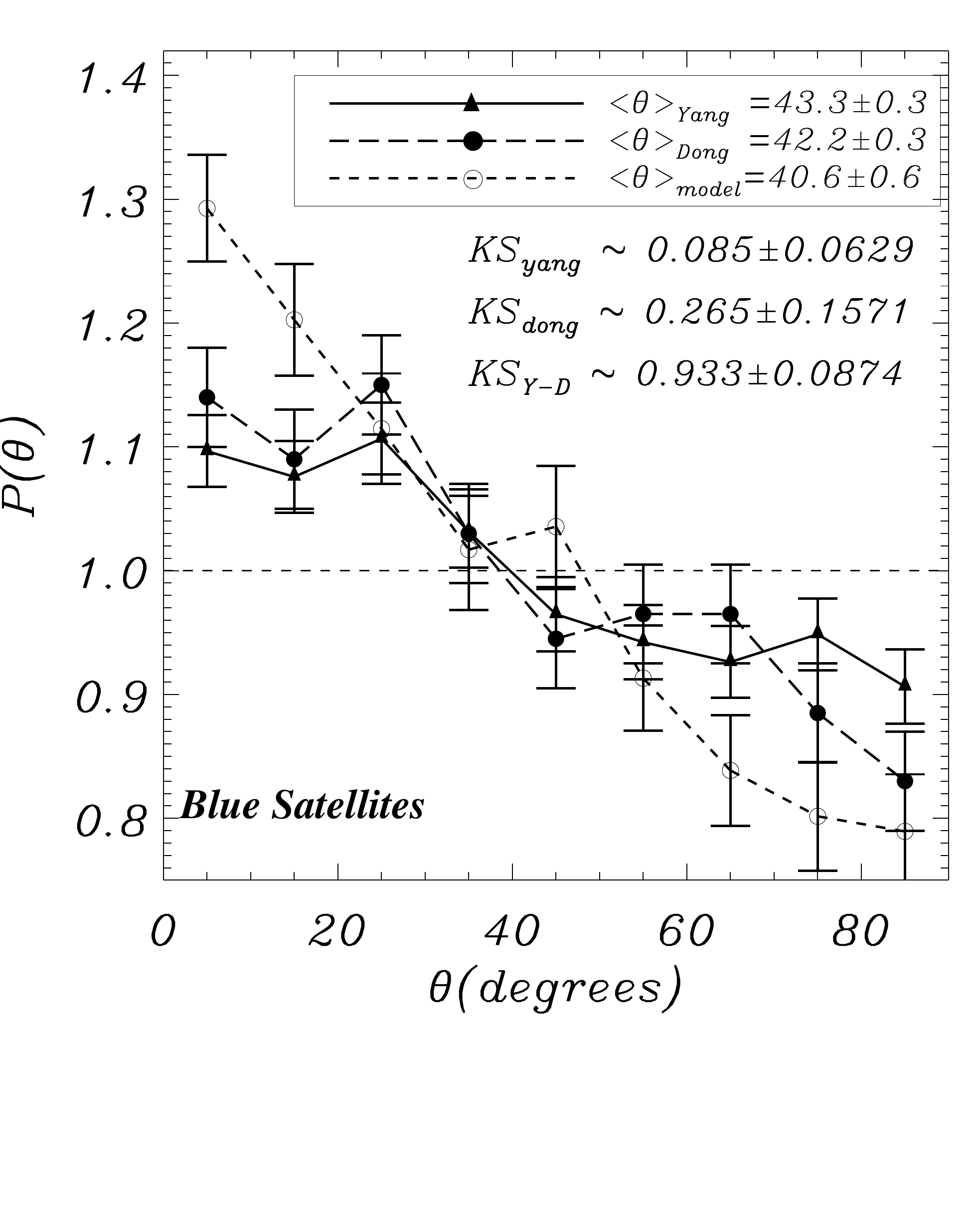}
  \vspace{-0.5cm}
  \vspace{-0.5cm}
  \includegraphics[width=0.45\textwidth, height=0.4\textheight]{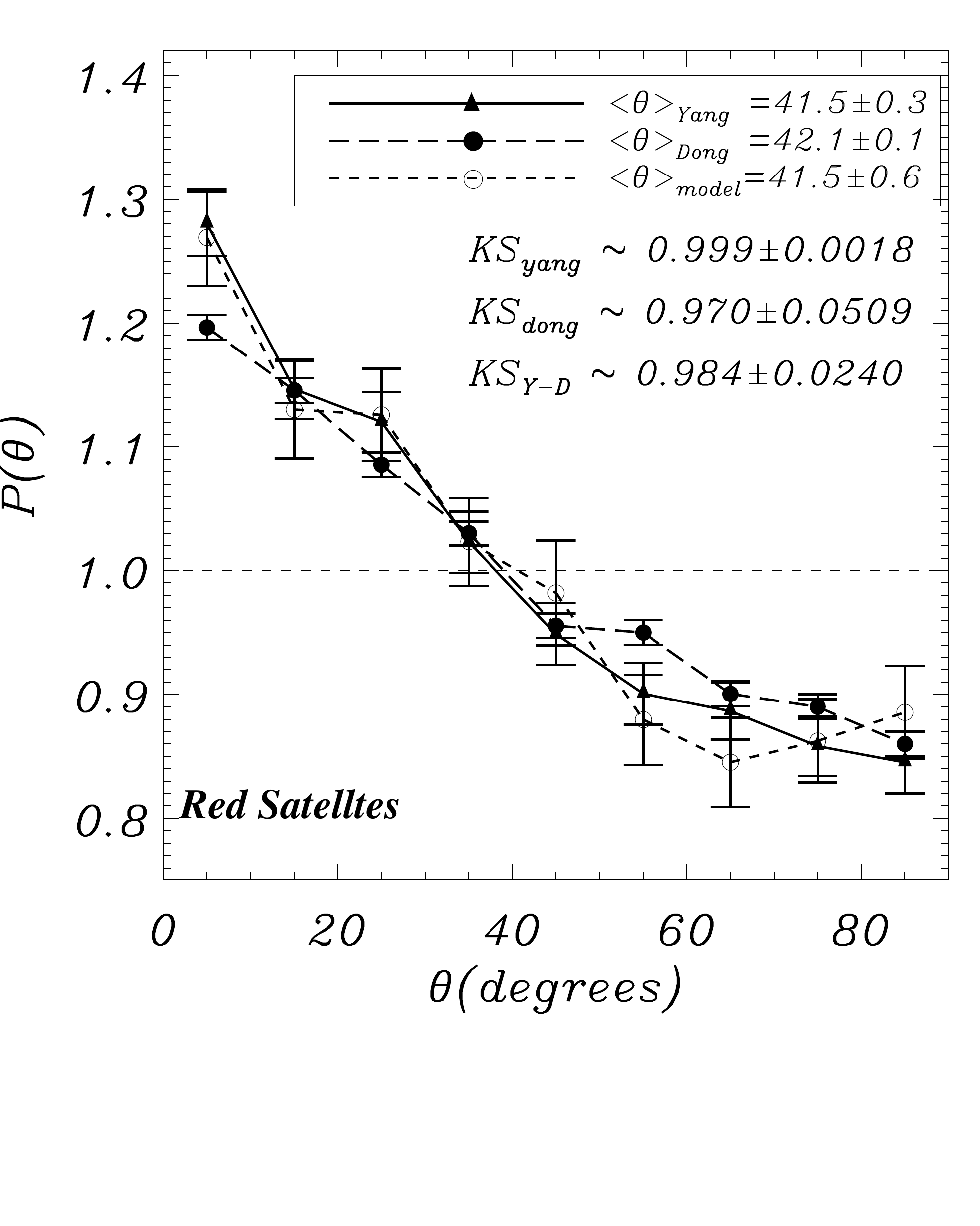}
  \vspace{-0.5cm}
  \vspace{-0.5cm}
  \includegraphics[width=0.45\textwidth, height=0.4\textheight]{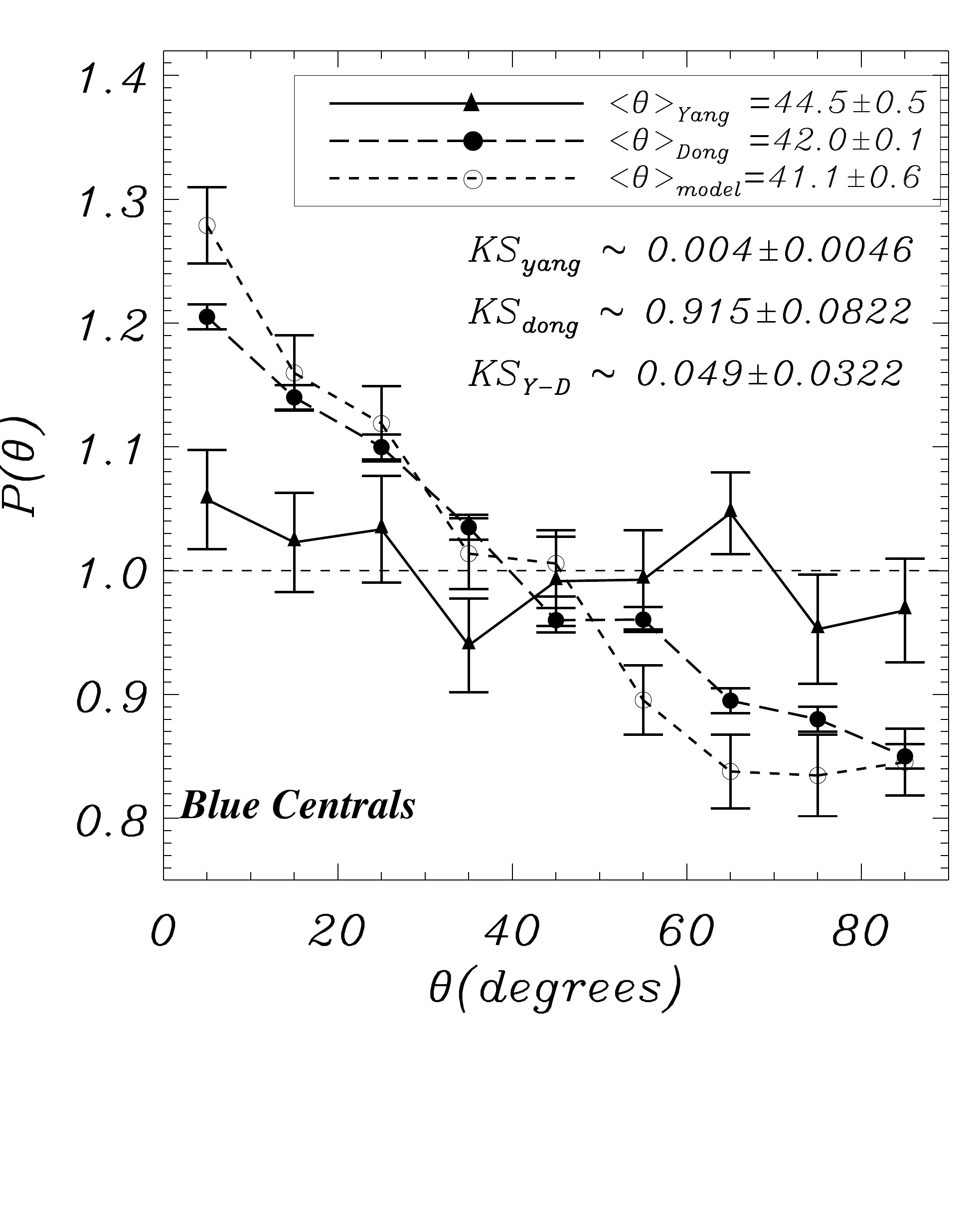}
  \vspace{-0.7cm}
  \includegraphics[width=0.45\textwidth, height=0.4\textheight]{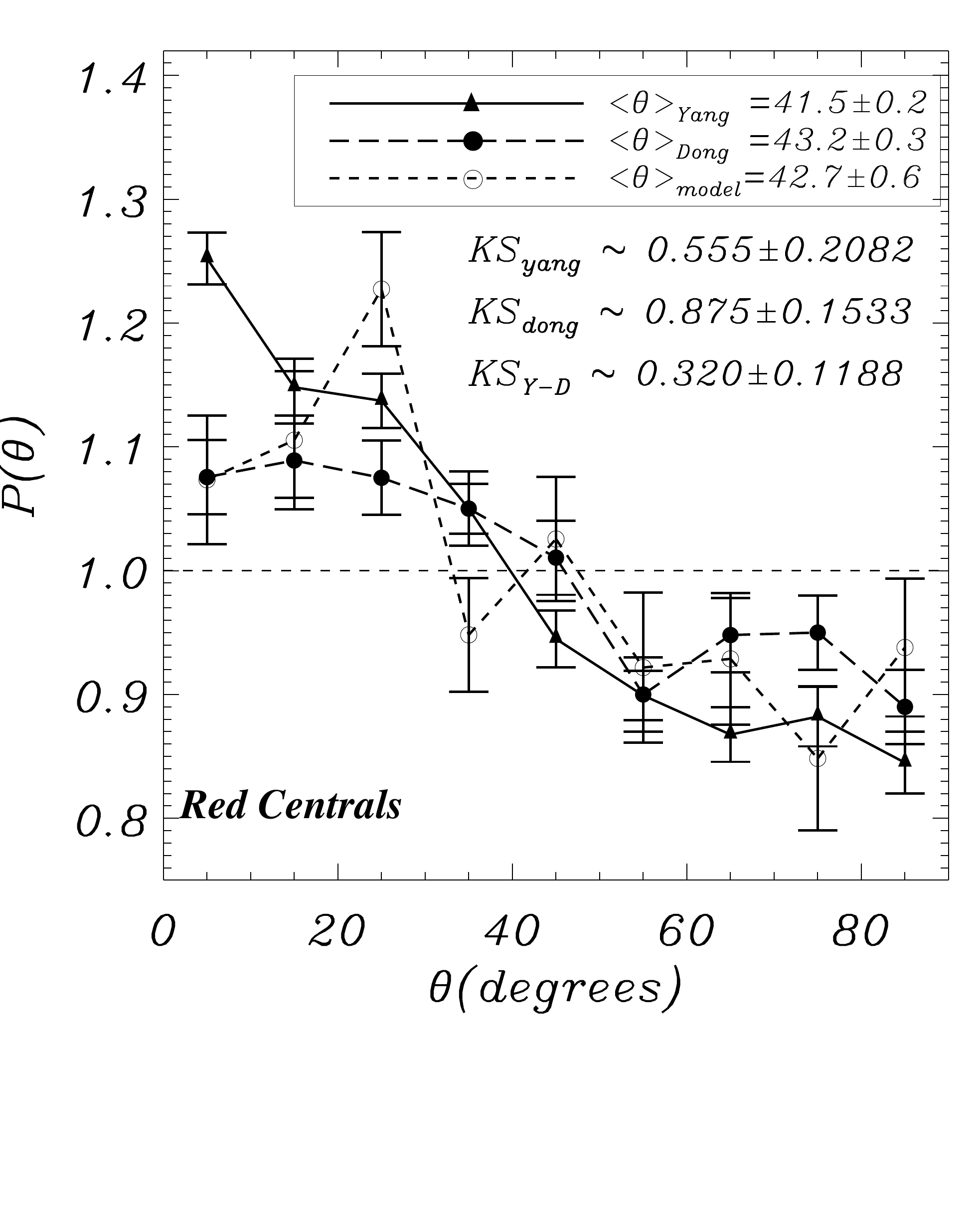}
  \vspace{-0.7cm}
    \caption{Predicted alignment of satellite galaxies at $z\sim0$ of different types of galaxies as indicated by the texts (short dashed line with blank circle), observational alignment (solid line with solid triangle) and previous theoretical prediction (long dashed line with solid circle). 
    The average angles $<\theta>$ are shown in the legends. 
    We show the standard deviation of the average angles.
    The statistical error is Poisson error.
    The $KS_{Yang}$ and $KS_{Dong}$ are the KS probability of drawing our prediction from the results in Y06 and in D14 for each subsample, respectively.
    The $KS_{Y-D}$ is the KS probability between the results in Y06 and in D14.
    Galaxy sample is combined with those on three projected planes.}
    \label{figure_probability_function_type}
\end{figure*}
%-------------------------%
%--satellite spatial distribution--%
\subsection{satellite alignment}\label{subsec:PF}
\par
For comparison with results of SDSS data at redshift $z\sim0$, we further apply the SDSS selection limits, $m_r<22.2 \ \rm magnitude$, and dark matter halos mass lower limit in Y06, $M>10^{12}h^{-1}M_{\odot}$, to obtain the comparative galaxy sample.
We also divide the comparative galaxies into four subsample, $Blue \ Satellites$, $Red \ Satellites$, $Blue \ Centrals$, $Red \ Centrals$, to predict the galaxy alignment and compare with previous works.
It is worth noting that $M>10^{12}h^{-1}M_{\odot}$ cut is applied for the purposes of comparison with the results in Y06 rather than accuracy (since it is noted in Section~\ref{sec:introduction} that this cut might lead to incompleteness issues).
\par
Figure~\ref{figure_probability_function_all} and Figure~\ref{figure_probability_function_type} show the predicted alignment of satellite galaxies comparing with previous studies. 
Galaxy sample is combined with those on three projected planes to increase the galaxy number, and reduce the bias.
$\theta \ (0^{\circ}\le\theta\le90^{\circ})$ is the angle on
the projection plane between the direction of satellites relative to the central galaxy and the major axis of their host central.
Satellite alignment strength is represented by the distribution probability function of $\theta$, $P(\theta)=N(\theta)/\langle N_{R}(\theta)\rangle$, where $N(\theta)$ is count of satellite in angular position bin $\theta$, while $\langle N_{R}(\theta)\rangle$ is the average count of satellites in the same bin $\theta$ from 100 groups of random samples, in which we randomize the orientation of all central galaxies.
The statistical error is Poisson error.
This calculation is same with that in Y06.
We compare the alignment signal with previous works by the Kolmogorov-Smirnov ($KS$) test. 
 We randomly value the $P(\theta)$ within the error margin in each bin, and repeat the KS computation 100 times.
The KS probability is the average of KS computations.
The error is the standard deviation of average KS probability.
The distributions are more similar, while the KS probability is much closer to 1 ($KS<0.2$ for two distinct distributions).
\par
As shown in Figure~\ref{figure_probability_function_all}, it is found that the probability function of our prediction roughly agrees with previous observation study and theory work, however our prediction of alignment signal looks slightly stronger than that in D14, utilizing the same simulation but with different galaxy identification. 
The KS probability of drawing ours from the results of Y06 and D14 is $0.963\pm0.0444$ and $0.765\pm0.0420$ respectively.
The $KS_{Y-D}$ is $0.999\pm0.0003$.
The interpretation is that our mock central galaxies have smaller radii and locate in the inner region of host halo, where stellar components express stronger alignment effect. 
This is in contrast with prediction of \cite{Wang2014a} where host halos were used to shape central galaxies.
 We can find from the comparison between those three KS tests  in Figure~\ref{figure_probability_function_all} that the alignment in our study and D14 both close to the observational result, but they are far from each other, for the significance levels.
We check the cumulative distribution functions (CDF) of those three alignments, and find that the CDF in D14 is lower the that in Y06, while the CDF in ours is higher than that in Y06. 
It is implied that the alignment is weaker than our result, but stronger than the result in D14.
Those three alignments are same, at significance levels.
%--------------------%
%--Figure_R_distribution--%
\begin{figure*}
  \centering
  \includegraphics[width=0.33\textwidth]{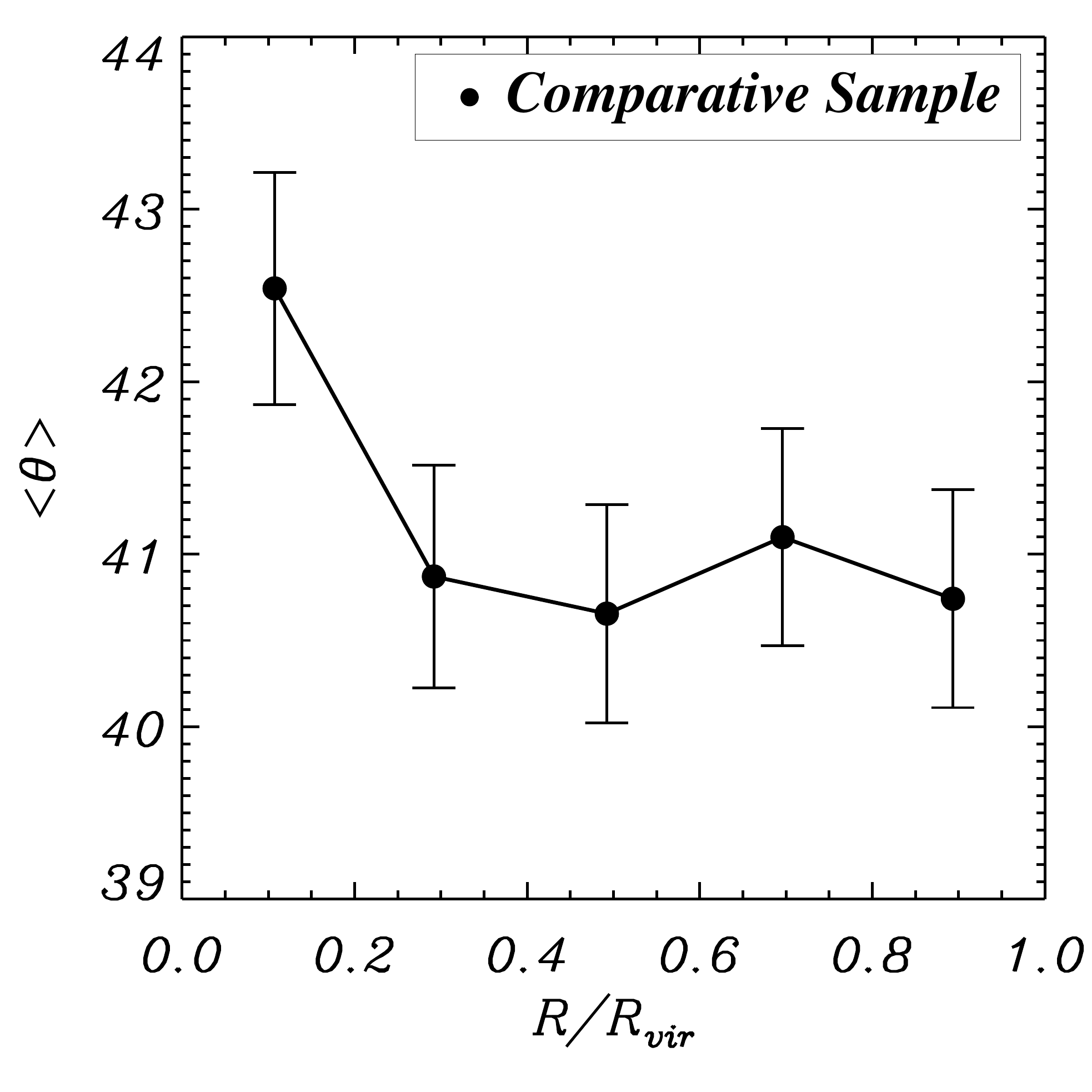}
  \includegraphics[width=0.33\textwidth]{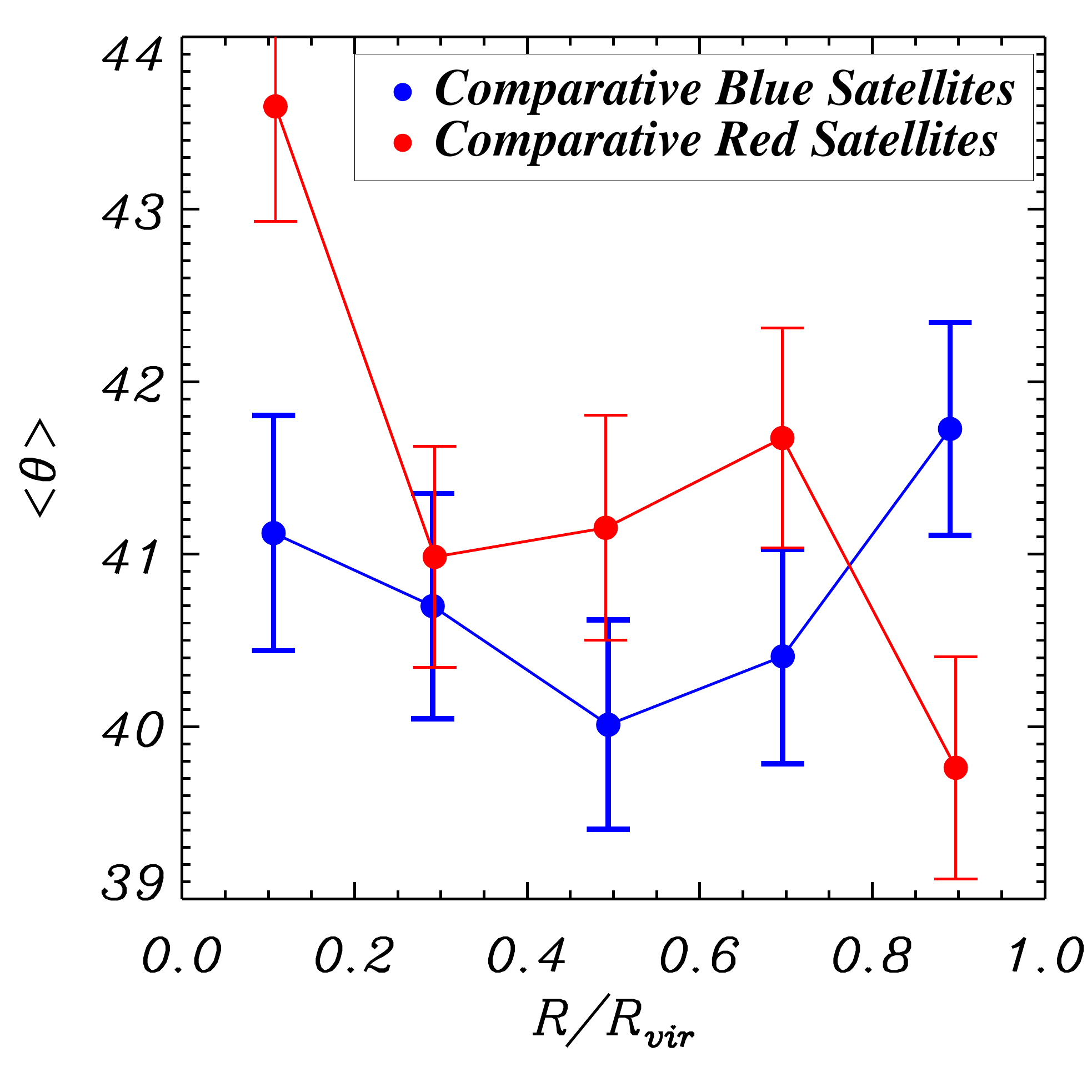}
  \includegraphics[width=0.33\textwidth]{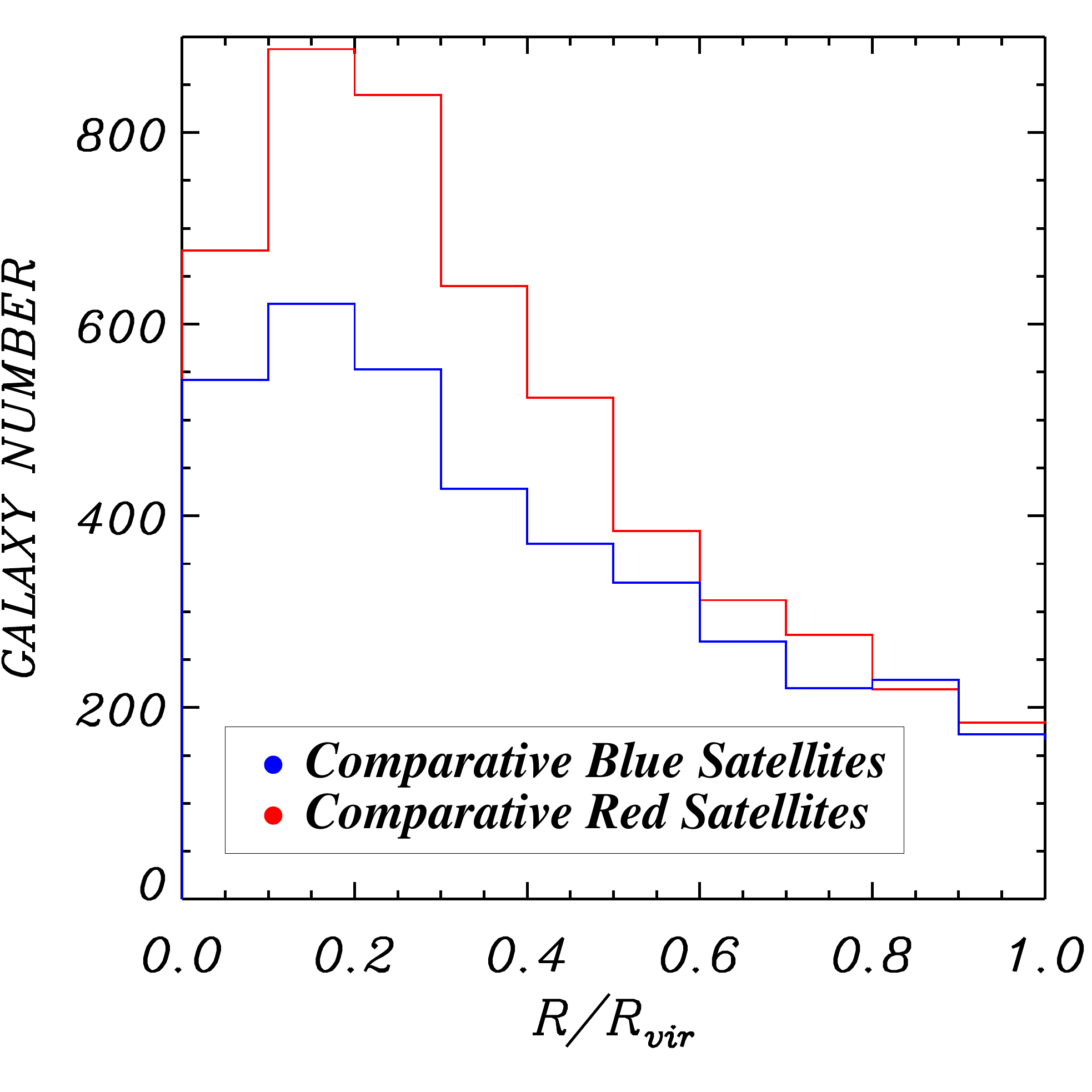}
    \caption{Dependence of average distribution angle on radius from the halo center normalized by halo viral radius at redshift $z\sim0$.
     Galaxy sample is combined with those on three projected planes in the left panel. 
     We divide galaxies into blue  (blue line) and red satellite (red line) subsamples, and plot the radial dependence in the middle panel and galaxy number density profile in  right panels, respectively.
     The error bar is the  standard deviation of average angles of satellites in each radius bin.
     }
    \label{figure_R_distribution}
\end{figure*}
%-----------------------------%
%--Figure_Para_R_distribution------%
\begin{figure*}
  \centering
  \includegraphics[width=0.4\textwidth]{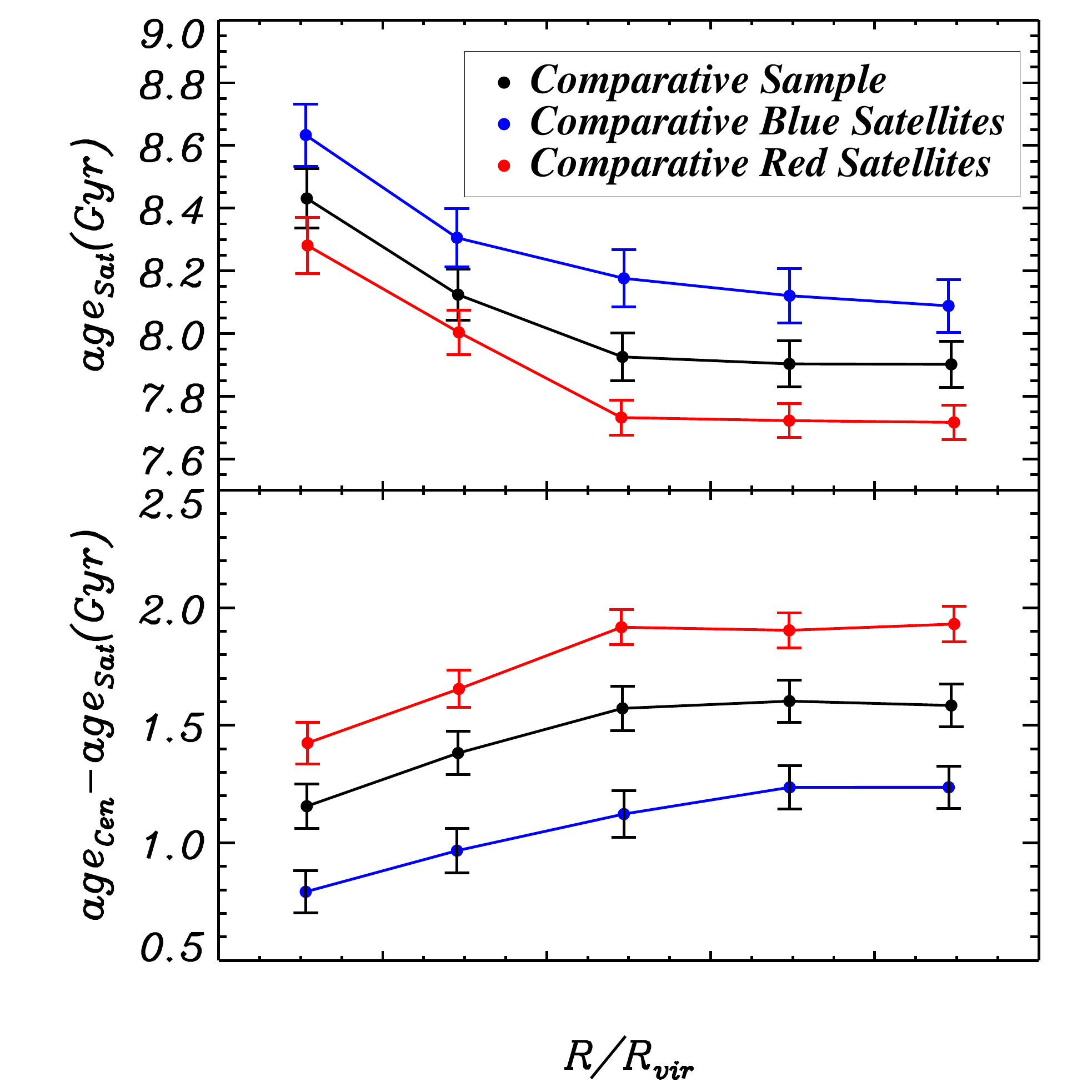}
    \includegraphics[width=0.4\textwidth]{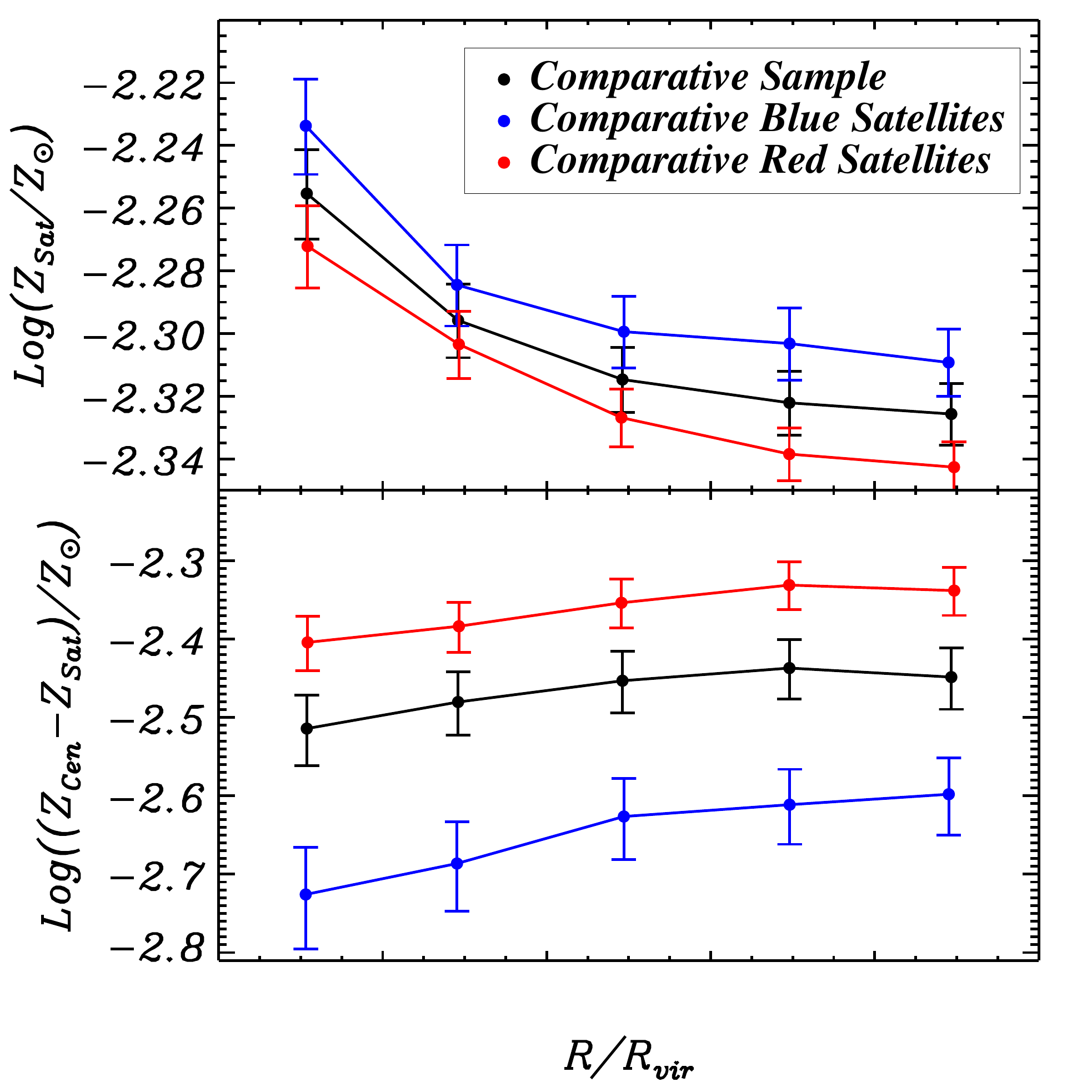}
   \caption{Dependence of satellite age and metallicity on radius. Top panel: radial distribution of satellites age and metallicity; Bottom panel: radial distribution of deviation of age and metallicity between central and its satellites. Black, red and blue lines are results of full, red and blue galaxy samples, respectively.
   The error bar is the standard deviation of average age and metallicity of satellites in each radius bin.
    }
    \label{figure_Para_R_distribution}
\end{figure*}
%-----------------%
\par
We divide the comparative sample into different galaxy types to predict the galaxy alignment and compare with previous works, as shown in Figure~\ref{figure_probability_function_type}.
For red subsamples, our result is comparable with that of previous works.
The KS probabilities of red subsamples are all larger than $0.3$.
Comparing with results in D14, the difference is obvious for blue subsamples.
For blue satellites, their number density in D14 increases with halo radius, contrary to the observation \cite[e.g.,][]{Wang2014b}.
The vast majority of blue satellites in the inner halos are missing from the sample in D14 (As discussed in Section~\ref{subsec:SRD}).
Because blue centrals are often non-spherical and actively star-forming, their dynamics and alignment are more strongly influenced by feedback processes than for red centrals.
When comparing with galaxy distribution in D14, we find that the misalignment of major axes of blue centrals between inner region and outer region is larger than that of red centrals, which causes a larger difference for blue centrals than for the red subsample of galaxies.
Because of the relative small sample causing by the SDSS selection limits, the alignment signal of red centrals shows a larger scatter than others.
\par
Comparing with results in Y06, the alignment signals of our blue subsamples are much stronger.
The prediction of alignment signals of the four subsamples looks similar to that in \cite{Kang2007} (Figure 3 in their paper).
However, because the simulation we used is lack of AGN feedback, the  centrals are too blue due to over cooling.
The color dependence need further investigation.
Finally, the difference between three kinds of KS value in each panel of Figure~\ref{figure_probability_function_type} implies that the alignment signal of red galaxies in our study is closer to the result in observation.
On the other hand, the blue galaxies in our study show much stronger alignment signals than observational results.

%---------------------------%
%--satellite radial distribution--%
\subsection{satellite radial distribution}\label{subsec:SRD}
%--Figure_evolution--%
\begin{figure*}
  \centering
  \includegraphics[width=0.33\textwidth, height=6cm]{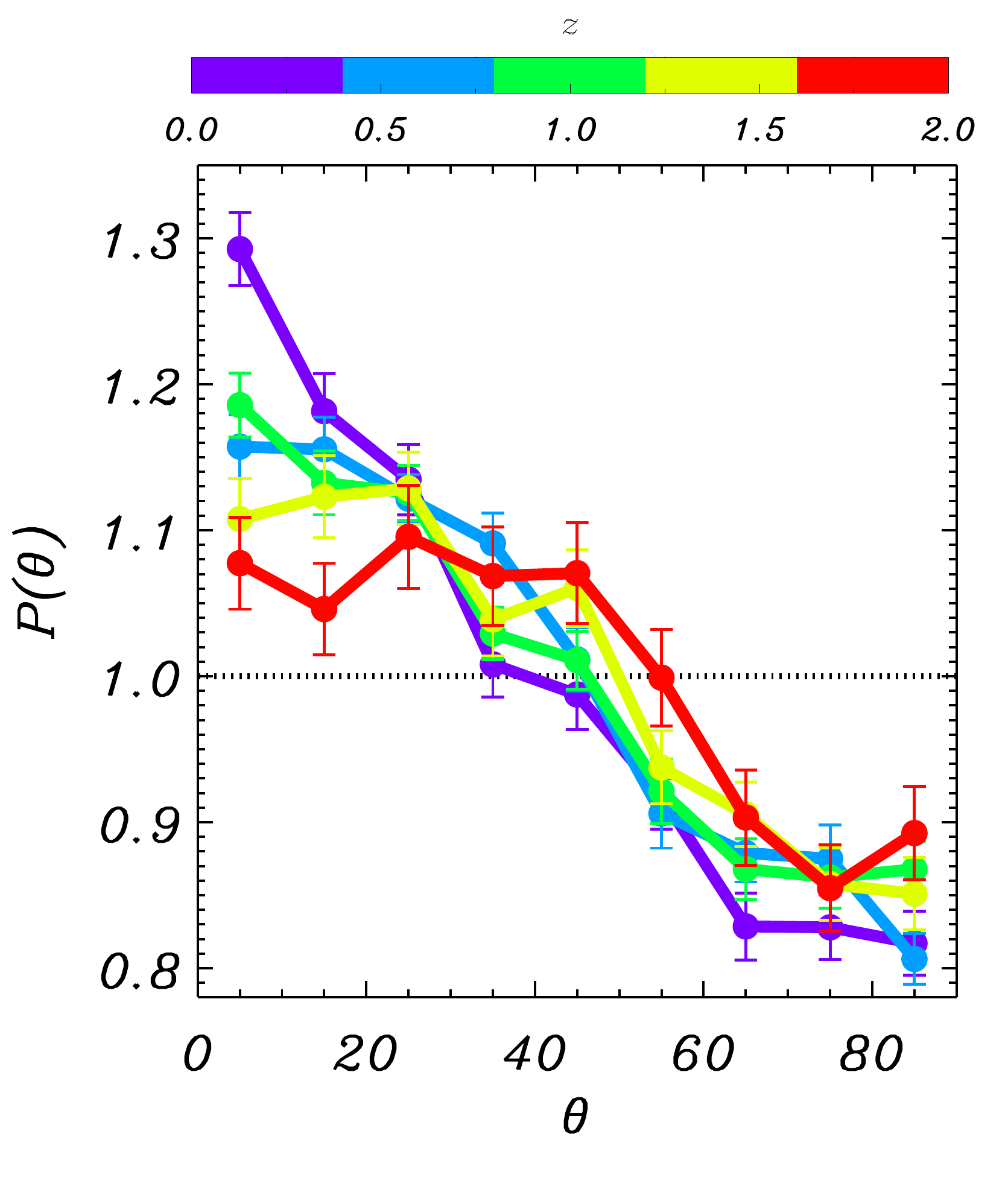}
  \includegraphics[width=0.33\textwidth, height=6cm]{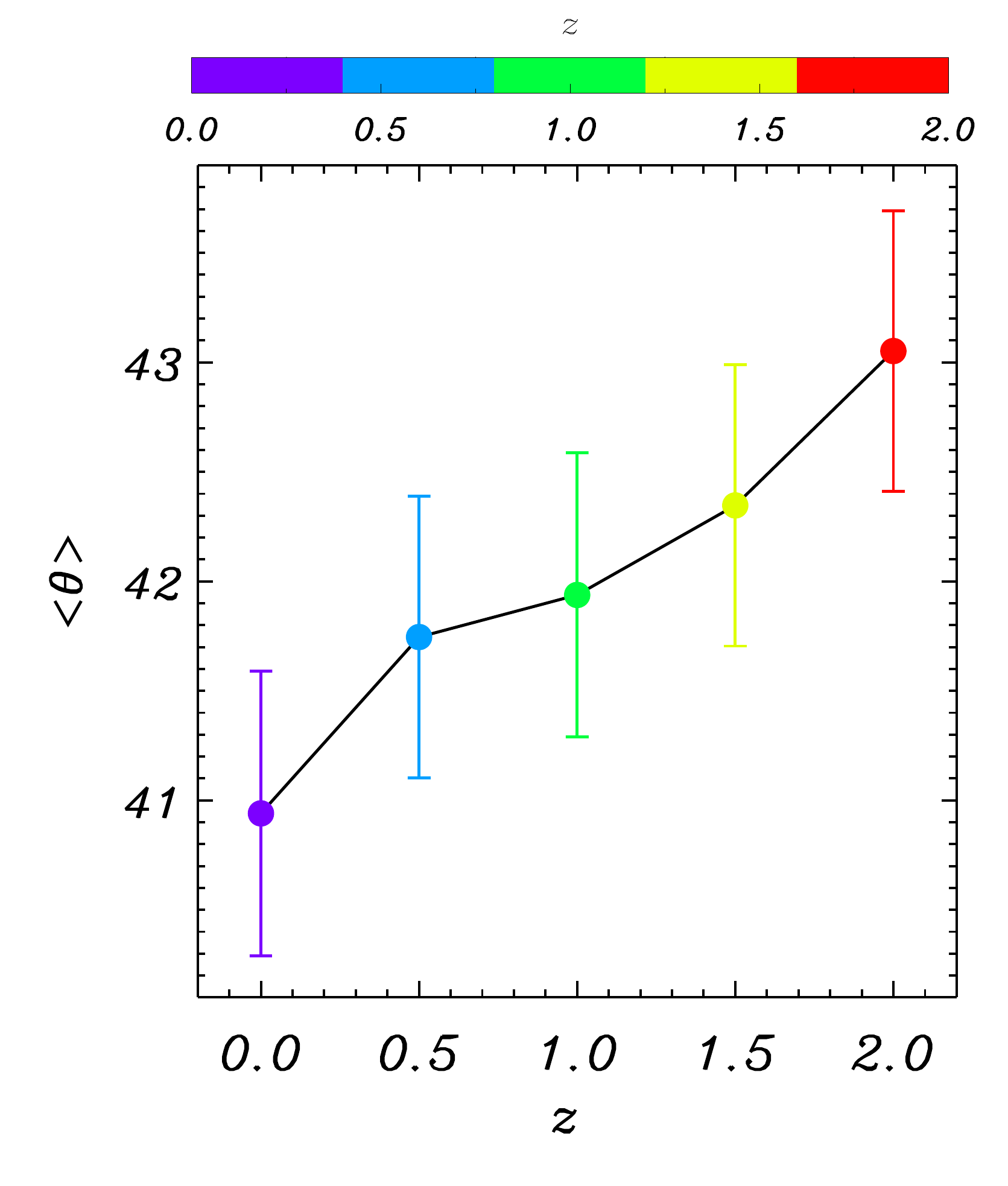}
  \includegraphics[width=0.33\textwidth, height=6cm]{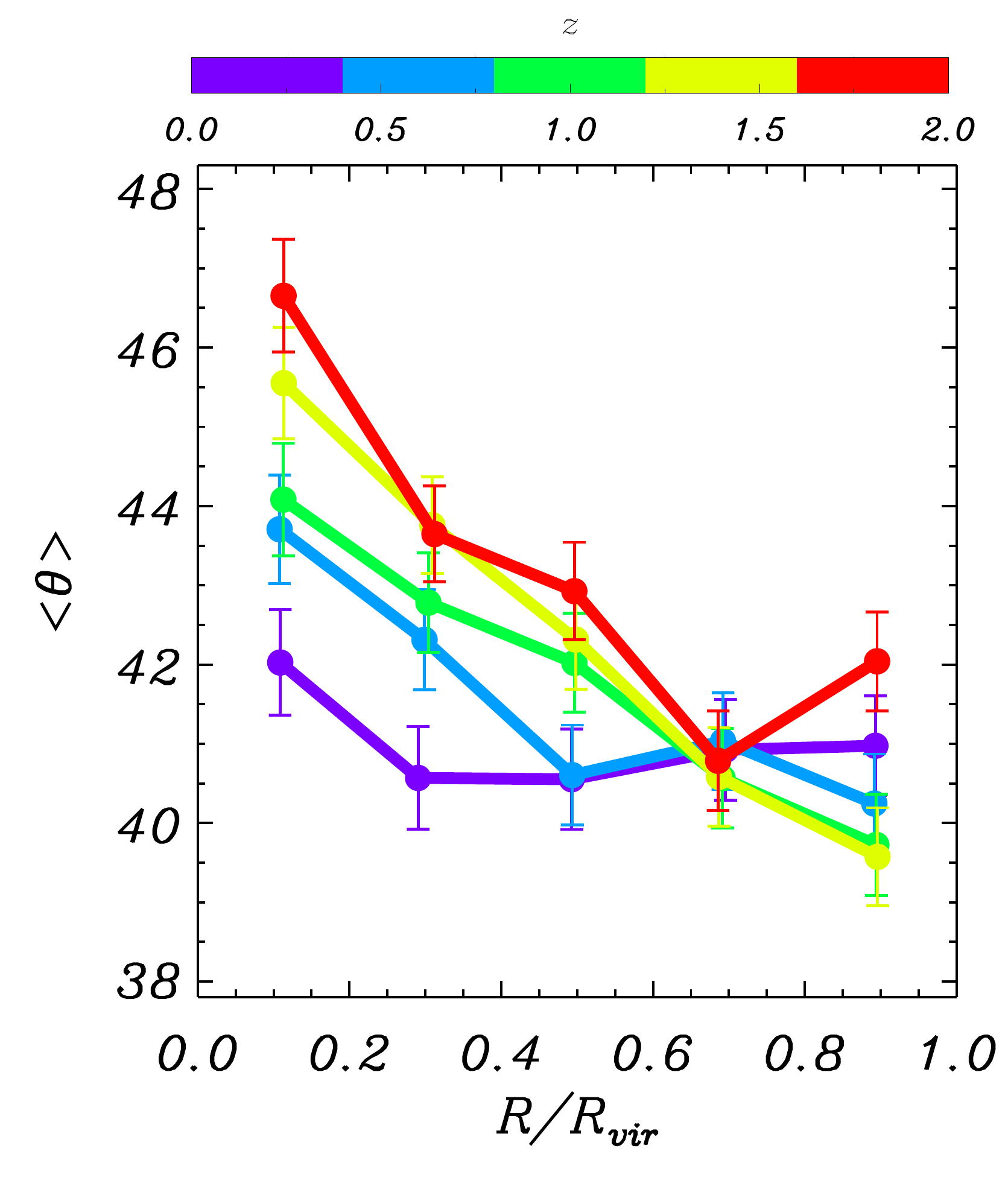}
    \caption{Left panel shows predicted alignment signals of satellites at different redshifts. 
    The statistical error is Poisson error.
    Middle panel is the dependence of average distribution angle on redshift. 
    Right panel shows the relation between average angle and radius of halos at different redshifts. 
   The error bars in middle and right panel are the standard deviation of average angles of satellites in each redshift and in each radius bin, respectively.
    The redshifts are $z=0, 0.5, 1.0, 1.5, 2.0$, as the color bar shows.
    The satellites are included by the fiducial samples.
    }
    \label{figure_evolution}
\end{figure*}
%-----------------%
%--table_KS_prob--%
\tabletypesize{\footnotesize} \tabcolsep=0.05cm
\begin{deluxetable}{lcccccccc} \tablecolumns{10}
\tablewidth{0pt}
\tablecaption{Kolmogorov-Smirnov (KS) probabilities at different redshifts.}
\tablehead{\colhead{ \ \ \ \ \ \ Sample redshift\ \ \ \ \ \ \ \ \ \ } & \colhead{ \ \ \ \ \ \ KS probability \ \ \ \ \ \ \ \ \ \ \ } }
\startdata
\\  \ \ \ \ \ \ \ \ \ \ \ \  0.0 &  $3.590\times10^{-4}$      
\\  \ \ \ \ \ \ \ \ \ \ \ \  0.5 &  $1.290\times10^{-3}$      
\\  \ \ \ \ \ \ \ \ \ \ \ \  1.0  &  $4.927\times10^{-3}$     
\\  \ \ \ \ \ \ \ \ \ \ \ \  1.5  &  $2.949\times10^{-2}$      
\\  \ \ \ \ \ \ \ \ \ \ \ \  2.0 &   $1.230\times10^{-1}$      
\enddata
\tablecomments{ The probabilities are for the angular distribution of the satellites to be drawn from an isotropic distribution.
}
\label{tab:KS_prob}
\end{deluxetable}
%------------------%
Left panel of Figure~\ref{figure_R_distribution} plots the radial dependence of the average position angle of satellites.
The distribution is in contrast with previous studies \cite[e.g.,][]{Dong2014,Wang2018}.
We check radial distributions on three projected planes, and find that there are similar distributions on three projected planes, but with big scatters.
In the middle panel of Figure~\ref{figure_R_distribution}, it is found that the radial distributions of blue and red satellites are hugely different.
In the inner region, blue satellites are more aligned with central galaxies than red satellites.
And dependence of red satellites on dark matter halo radius is opposite to that of blue galaxies, particularly in the outer region. 
\par
We check the galaxy radial number density profile, and find that the number of red satellite is much larger than that of blues, particularly in the inner halos.
This density profile is more agreeable with observational profile \cite[]{Wang2014b} than that in D14, as shown in the right panel of Figure~\ref{figure_R_distribution}.
Comparing with galaxy sample defined by the traditional substructure finder in D14, our sample includes more galaxies in the most inner region of dark matter halos.
Those extra galaxies are close to their host centrals, while they are commonly treated as part of centrals by traditional substructure finders.
As shown in the middle panel of Figure~\ref{figure_R_distribution}, the average distribution angle of inner galaxies ($<0.5R/R_{vir}$) decreases with increasing radius.
Lacking satellites in the inner region naturally causes the contrast of galaxy radial distribution between D14 and this work.
\par
It has been proved that satellites distribution is strongly dependent on galaxy properties \cite[e.g.,][]{Yang2006, Brainerd&Yamamoto2019} and the satellite alignment signal strongly correlates with satellite metallicity \cite[e.g.,][]{Dong2014}.
We study the color-metallicity-age relation, and found that satellites with higher metallicity have bigger age, but for a given color, satellite age shows a huge distribution.
It implies that the galaxy alignment dependence on metallicity is the reflection of age relation. 
\par
So as to discover the physical mechanism of satellites radial distribution, we plot the dependence of age and metallicity of satellites on radius, as shown in Figure~\ref{figure_Para_R_distribution}.
The age and metallicity of galaxies are calculated by the average of age and metallicity of total components included by the galaxies.
The age and metallicity in each bin of $\Delta(R/R_{vir})$ are defined by the average age and metallicity of satellites located in region from $R/R_{vir}$ to $R/R_{vir}+\Delta(R/R_{vir})$.
It is found that age and metallicity of satellites decreases with radius. 
Blue satellites are older and metal richer than red satellites within virial radii of dark matter halos. 
On the other hand, the age and metallicity deviation between central and its satellites increases with larger radius.
Compared with red satellites, blue satellites have their age and metallicity much closer to their central galaxies.
It seems that blue satellites are population close to centrals, but red satellites are galaxies accreted from nearby structure.
The dependence of age and metallicity of satellites on radius in Figure~\ref{figure_Para_R_distribution} imply that satellites located in inner halos are older, metal richer and much closer to their host centrals than those located in outer halos.
\par
The dependence of age and metallicity of satellites on halo radius illustrates that satellites located in inner halos and blue satellites are more likely formed at the same time as the central galaxy or early-merger remnants, while the distribution of those satellites seems to better trace the dark matter distribution \cite[e.g.,][]{Jing&Suto2002, Yang2006}. 
On the other hand,  satellites located in outer halos and red satellites are late-merger or accretion remnants, and their distribution are more asymmetrical \cite[e.g.,][]{Wang2005}.

%-------------------%
%%--redshift evolution--%
\subsection{redshift evolution}\label{subsec:evolution}

Figure~\ref{figure_evolution} shows the dependence of satellites spatial distribution on redshifts.
The left panel shows the predicted alignment signals of satellites at four different redshifts $z\sim0.0, 0.5, 1.0, 1.5, 2.0$.
In Table~\ref{tab:KS_prob}, we list the KS probabilities of the predicted alignments drawn from an isotropic distribution.
The KS probability is bigger at high redshifts than that at low redshifts.
Those results illustrate that satellites distribution signal is stronger at lower redshifts.
Middle panel represents that average angle of satellite distribution is larger at a higher redshift.
Those dependence of distributions on redshifts imply that satellites become more inhomogeneous with cluster evolution, which is consistent with the alignment signal in MassiveBlack II simulation \cite[e.g.,][]{Bhowmick2020} and expected in \cite{Wang2010}.
There have been some results that are inconsistent with ours, for example, in \cite{Donoso2006}, alignment signal at z=0.5 in SDSS DR4 is similar to the one in local universe. 
However, their average distribution angle of satellites is $~44^{\circ}$, which is a weaker alignment signal than that in Y06 ($41.3^{\circ}$).
\par
Right panel of Figure~\ref{figure_evolution} shows the relation between radial distribution and redshift. 
The deviation of average angle between inner region and outer region increases with redshifts. 
The average angle of satellites spatial distribution in the inner region decreases, meanwhile that in the outer region increases, with cluster evolution.
\par
The radial dependence of satellite distribution is getting weaker with time, but its original information is impressed in the dark matter halos and is not erased or reversed by the halo evolution.
We test the evolution of radial distribution for blue and red subsamples.
It is found that the radial distribution of blue satellites appears similar to that of red satellites at high redshifts ($z=0.5,1.0,1.5,2.0$), and differs only at larger radius with smaller average angle.
In the inner part of host halo, both red and blue satellites in early universe show higher average distribution angles than those in local universe.
It is implied that the galaxies are far away from the direction of centrals' major axis due to post-infall evolution, but this influence decreases with time.
%--------------------------%
%--Conclusions and Discussion--%
\section{Conclusions and Discussion}\label{sec:conclusions}

A wide variety of works has investigated the satellite spatial distribution.
Current observations and theoretical results consistently agree that satellites are preferentially distributed along the major axes of host centrals, but slightly different in terms of intensity of alignment signal. 
Many kinds of theoretical study attempted to reduce the discrepancy with observations, and achieved success on some degree.
\par
In this paper, we explore the satellite distribution using an observational mimic galaxy definition, which differs from traditional substructure defined algorithms in hydrodynamical simulations. 
Our method is based on the projected observational surface brightness profile of each FoF group to define mock galaxies, with observational parameters.
Using several surface brightness limits mock galaxy sample is reconstructed. 
This method avoids multiple galaxies connected together and narrow galaxy region for single surface brightness limit, as shown in Figure~\ref{figure_Luminosity_profile} and Figure~\ref{figure_Luminosity_profile_reconstruction}.
In this way, the density profile of galaxy is more consistent with that in observations, as shown in the right panel of Figure~\ref{figure_R_distribution}.
\par
We study the alignment of satellites comparing with previous predictions, dependence of average angle of satellite distribution on viral radii of dark matter halos.
We also study the redshift evolution of satellite distribution, which was rarely discussed in the literatures. 
The results we obtained are summarized as follows:
\begin{itemize}
\item[1.]{%satellite spatial distribution
The alignment of our reconstructed mock galaxies are approximately agree with previous predictions.
There are two discrepancies: 1) overall probability functions show slightly stronger alignment signals than those of previous predictions; 2) the alignments of blue subsamples exhibit obviously stronger signals than observations.
}
\item[2.]{%satellite radial distribution
Alignment strength of satellites is  anti-correlated correlated with the virial radii of dark matter halos in our work.
It shows similar results for three projected images.
We subsample the galaxy sample into red and blue galaxies, and find that blue satellites are more aligned with major axes of centrals residing in the inner regions.
And the radial distribution of satellites age and metallicity decreases with radius.
The age and metallicity of blue satellites are much closer to that of host centrals comparing with red satellites.
}
\item[3.]{%redshift evolution
The dependence of satellites spatial distribution on redshifts is small but exists.
Generally, the lower redshift, the stronger alignment signal, and the flatter radial distribution curve.
The strength of alignment signal in our predictions is in broad agreement with observational results. More observation data is need to calculate satellite alignment signals at high redshifts. 
}
\end{itemize}
\par
In summary, comparing with traditional galaxy finder, our predicted satellite alignment is slightly stronger, particularly for blue subsamples. 
Defined galaxies (especially centrals) by the traditional galaxy finder include much outer components, and galaxy number density is underestimated in the inner region of dark matter halos \cite[e.g.,][]{Klypin1999, Liu2010, Onions2012}.
We illustrate that our predicted probability function of satellite distribution are still stronger than observational results.
Considering a lack of AGN feedback in our simulation, we state that a complete galaxy formation model, e.g., including AGN feedback \cite[e.g.,][]{Scannapieco2012, Sembolini2016, Cui2016} is important to reproduce the galaxy structure, or small scale structure of universe.
Overall, satellite spatial distribution is dominated by red galaxies, but sensitive to blues. 
Although, our simulation can not perfectly reproduce the satellite distribution, the comparison between prediction of simulations and observations is much fairer than that in previous studies.
The difference between our predictions and the satellite alignment signal measured by Y06 is more reasonable than in previous studies due to our reconstruction method, which has been designed to mimic the observations of ICL \cite[]{Tang2018}.
\par
Our studies also imply that satellites are fallen into the halos from the direction of centrals' major axis, while the post-infall evolution has strong influence on the alignment signal.
The post-infall influence decreases the strength of alignment signal, causing the dependence of satellites average angle on radii of host halos in our study.
 However, the decreasing influence of post-infall evolution is getting weaker for satellites with later infall times.
Alignment signal of satellites is due to post-infall evolution and infall of galaxies along the nearby large scale structure.
This results that satellites are distributed more asymmetrically in local universe.
\par
Comparing the radial distribution, it is found that missing of galaxies in the inner host halos in D14, is the main reason of the difference between radial distribution in our study and in D14.
The galaxy sample in Y06 and \cite{Wang2018} is obtained by a halo-based group finder \cite[]{Yang2005}, which is based the FoF algorithm.
We will apply our galaxy finder to SDSS data to study the difference between galaxy sample in Illustris simulations \cite[]{Vogelsberger2014} and the sample in Y06 and \cite{Wang2018} in future work.
Considering the incomplete galaxy formation model in our simulation code, we plan to re-examine the satellite radial distribution using results of Illustris simulations \cite[]{Vogelsberger2014} and IllustrisTNG simulations \cite[]{Pillepich2018,Pillepich2019}, which includes a much more complete galaxy formation model. 
The Illustris simulation has been used to analyze satellite distribution by \cite{Brainerd&Yamamoto2019}, and it is shown that the median angle of three dimensional satellites radial distribution decreases with radius within host dark matter halo.
\par
In our studies, the dependence of satellite age and metallicity on halo radius can interpret the anti-correlation between alignment strength and halo radius. 
Satellites located in inner halos and blue satellites have more similar age with host centrals than satellites located in outer halos and red satellites, meaning that they are more likely concentrations by gravitational potential located at primordial density peaks within dark matter halos, according to the hierarchical clustering formation theory, or early-merger remnants.
Therefore, the distribution of those kinds of satellites traces the dark matter distribution, which are strongly aligned in the inner part of host halos \cite[e.g.,][]{Jing&Suto2002, Kang2005b, Wang2014a}.
On the other hand, satellites located in outer halos and red satellites are more likely late asymmetrical merged or accretion remnants from surrounding environment or large-scale structure (LSS) \cite[e.g.,][]{Wang2005, Wang&Kang2018}.
\par
Considering the dependence of alignment of satellites on LSS \cite[e.g.,][]{Zhang2015,Wang2018}, in the dense environments of LSS, the alignment signal of satellites is strengthened from $Cluster$ (or $Knot$), $Filament$ to $Sheet$ (or $Wall$), which is similar to our results that alignment signal is reduced by the post-infall evolution.
Because the angular momentum of matter in the Knot's environment is chaotic, the trajectory of satellite galaxies after falling is more likely to deviate from the direction of centrals' major axis \cite[e.g.,][]{Zhang2019}.
\par
The redshift evolution illustrates that current galaxy distribution in dark matter halos is less homogeneous and more aligned with major axes of host centrals than early universe.
In another word, in early universe, the galaxy distribution is almost symmetric.
The evolution of galaxy distribution exhibits two kinds of galaxy dynamical history, implying two physical effects on galaxies, in the inner and outer halo region, respectively.
 With evolution, the outer galaxies are gradually located in a wider range caused by the halo gravitational potential.
 Meanwhile, the inner galaxies distribute more preferentially aligned with the major axes of central galaxies influenced by the galaxy dynamical interaction and central astrophysical mechanism, which are major impact on the evolution of small scale cosmic structure \cite[e.g.,][]{Springel2005b, Mo2010}. 
This scenario also explains why satellites distribution today is more inhomogeneous.

%------------------%
%--acknowledgments--%
\acknowledgments

Acknowledgments. 
The authors thank the anonymous referee for useful suggestions. 
W.P.L. acknowledges support from the National Key Program for Science and Technology Research and Development (2017YFB0203300), the National Key Basic Research Program of China (No. 2015CB857001) and the NSFC grant (No.11473053).
W.Y acknowledge supports by the NSFC projects (No.11643005).
The simulations were run in the Shanghai Supercomputer Center and the data analysis was performed on the supercomputing platform of Shanghai Astronomical Observatory and School of Physics and Astronomy, Sun Yat-sen University. 
%--------------%
%--bibliography--%
\bibliography{SateDistr}
%--------------%
\end{CJK*}
\end{document}